\documentclass[11pt]{article}

\usepackage[margin=1in]{geometry}
\usepackage{amsmath,amssymb,amsthm,mathtools,bm}
\usepackage{fix-cm}
\usepackage{booktabs,multirow}
\usepackage{caption}
\DeclareCaptionLabelSeparator{simulationbar}{\quad\textbar\quad}
\usepackage{enumitem}
\usepackage{natbib}
\setcitestyle{authoryear,round}
\usepackage{xcolor}
\usepackage{hyperref}
\usepackage{microtype}
\usepackage{graphicx}
\usepackage{pdflscape,setspace}
\usepackage[ruled]{algorithm}
\usepackage[noend]{algpseudocode}
\algrenewcommand{\algorithmicrequire}{\textbf{Input:}}
\algrenewcommand{\algorithmicensure}{\textbf{Output:}}
\algrenewcommand{\alglinenumber}[1]{\scriptsize\color{gray}#1:}
\newcommand{\bI}{\bm I}
\newcommand{\bB}{\bm B}
\newcommand{\bu}{\bm u}
\newcommand{\bU}{\bm U}
\newcommand{\bW}{\bm W}
\newcommand{\br}{\bm r}

\newcommand{\bY}{\bm Y}
\newcommand{\bX}{\bm X}
\newcommand{\bZ}{\bm Z}
\newcommand{\bt}{\bm t}

\newcommand{\bomega}{\bm\omega}
\newcommand{\boldeta}{\bm\eta}
\newcommand{\btheta}{\bm\theta}
\newcommand{\bzeta}{\bm\zeta}
\newcommand{\bTheta}{\bm\Theta}
\newcommand{\bSigma}{\bm\Sigma}
\newcommand{\bPhi}{\bm\Phi}
\newcommand{\bLambda}{\bm\Lambda}
\newcommand{\bbeta}{\bm\beta}
\newcommand{\bgamma}{\bm\gamma}
\newcommand{\bzero}{\bm 0}
\newcommand{\calA}{\mathcal A}

\newcommand{\Gmax}{G_{\max}}
\newcommand{\Prb}{\mathbb P}
\newcommand{\E}{\mathbb E}
\newcommand{\I}{\mathbb I}

\newtheorem{theorem}{Theorem}

\title{\textbf{Unifying Data-Adaptive Longitudinal Treatment-Effect Subgroup Discovery and Pretreatment Membership Prediction}}
\author{Pu Zhang, Yuheng Li and Jiakun Jiang\\
Center for Statistics and Data Science, Beijing Normal University}
\date{\today}

\begin{document}
\maketitle

\begin{abstract}
Precision medicine seeks to match patients with treatments from which they are most likely to benefit. A key step toward this goal is to learn treatment-effect heterogeneity from longitudinal study data. This requires discovering subgroups with distinct treatment-effect trajectories and predicting subgroup membership for new patients from pretreatment information. Existing methods do not fully unify these tasks. Covariate-informed membership models typically prespecify the subgroup number or covariance structure, whereas data-adaptive discovery methods often require prediction in a separate stage.
We propose a data-adaptive semiparametric mixture-of-experts framework that unifies longitudinal treatment-effect subgroup discovery and pretreatment membership prediction. Subgroup-specific varying-coefficient functions characterize treatment-effect trajectories, while flexible covariance functions account for within-subject dependence. A covariate-dependent gating model links membership to pretreatment characteristics. A penalty on global component weights removes redundant subgroups, while separate gating coefficients allow patient-specific membership probabilities among retained subgroups. A blockwise generalized expectation--maximization algorithm estimates all model components under a single penalized observed-data objective. We establish consistent subgroup-number selection, convergence rates for the functional estimators, and asymptotic normality for the finite-dimensional parameters.
Simulations show reliable subgroup-number selection and more accurate conditional membership prediction than a two-step procedure. Applications to the SANAD epilepsy study and the COMPETE migraine trial identify distinct temporal treatment-effect patterns. 
\end{abstract}

\section{Introduction}
A central goal of precision medicine is to use pretreatment information to guide treatment selection for new patients. Conventional analyses often summarize treatment benefit through a population-average effect, which may obscure important differences across patients \citep{kent2018,lipkovich2017}. Subgroup analysis of longitudinal outcomes can characterize this heterogeneity by identifying groups with distinct treatment-effect trajectories. To translate outcome-defined subgroups to new patients, membership must be predictable from information available before treatment. This calls for a unified model that learns the subgroup number and treatment-effect trajectories from study outcomes while predicting membership from pretreatment covariates.

Longitudinal follow-up provides a richer basis for studying treatment-effect heterogeneity than a single endpoint. Repeated measurements reveal when treatment responses emerge, how they evolve, and whether they persist, and may also improve the precision of treatment comparisons \citep{collins2006,gibbons2010,schober2018}. These advantages have motivated a substantial literature on statistical methods for longitudinal treatment-response analysis. At the same time, longitudinal data introduce within-subject dependence and irregular observation times, while treatment-response trajectories may be nonlinear. In heterogeneous populations, both treatment-effect trajectories and within-subject covariance structures may vary across latent subgroups, requiring flexible subgroup-specific modeling of both features.

Existing methods fall into two broad, partly overlapping strands. 
Covariate-defined approaches construct treatment-sensitive subgroups 
from pretreatment characteristics. Threshold and change-plane models 
define subgroup boundaries through one or more covariates, with 
extensions for dropout, time-varying treatment effects, and floor or 
ceiling effects \citep{ge2020,wei2022,wei2023,ge2024}. Tree-based 
methods recursively partition the covariate space and accommodate 
longitudinal or multivariate outcomes \citep{loh2016,wei2020,xu2026}. 
Outcome-driven approaches instead infer latent heterogeneity from 
repeated responses. Random-effects mixtures and structured mixed-effects 
models identify subgroups with distinct treatment-response trajectories; 
some also model membership probabilities using pretreatment covariates 
\citep{xuhedeker2001,shenqu2020,wen2024}.

More recent work has pursued data-adaptive subgroup discovery and 
membership prediction, although often separately. \citet{zhang2026} 
proposed a semiparametric longitudinal treatment-effect model that 
selects the subgroup number and estimates subgroup-specific covariance 
functions. Related methods include sparse mixture regression for 
multivariate functional responses \citep{zhong2025} and fusion-based 
clustering of longitudinal profiles \citep{zhuqu2018}. For 
nonlongitudinal outcomes, \citet{shenhe2015} jointly identify a 
treatment-sensitive subgroup and construct predictive scores for 
membership under a structured mixture model. \citet{chen2026} considered 
a scalar outcome with a multivariate grouping vector, allowing 
coefficient-specific subgroup structures; their procedure first detects 
groups by regularization and then uses a support vector machine to 
recover covariate-space partitions for predicting future membership.

Taken together, these methods do not jointly address data-adaptive subgroup discovery and pretreatment membership prediction. Methods that incorporate pretreatment covariates often 
prespecify the subgroup number or within-subject covariance structure. The threshold and change-plane 
methods cited above, as well as \citet{shenqu2020}, assume two 
subgroups. \citet{wen2024} select the number of classes by comparing 
prespecified candidate models and retain a parametric model for 
within-subject dependence. In contrast, the data-adaptive longitudinal 
methods discussed above do not estimate a pretreatment membership model 
jointly with subgroup discovery. \citet{zhang2026}, for example, fit a 
membership model only after discovering the subgroups; in a 
nonlongitudinal setting, \citet{chen2026} similarly recover 
covariate-space partitions with a support vector machine after detecting 
coefficient groups. Because pretreatment covariates do not contribute to 
the discovery stage, these procedures treat estimated membership 
labels or responses as fixed during prediction, so discovery errors can
propagate to the prediction stage.

We address this gap with a data-adaptive semiparametric 
mixture-of-experts model that jointly discovers longitudinal 
treatment-effect subgroups and predicts membership from pretreatment 
covariates. Within each subgroup, varying-coefficient functions 
characterize the treatment-effect trajectory, while a flexible 
covariance function captures within-subject dependence. A 
covariate-dependent gating model relates subgroup membership to 
pretreatment characteristics. Beginning with a prespecified upper bound 
on the number of subgroups, we penalize global component weights to 
eliminate redundant groups, while separate gating coefficients allow 
membership probabilities among the retained groups to vary across 
patients. A blockwise generalized expectation--maximization algorithm 
jointly estimates all model components under a single penalized 
observed-data objective. For study participants, posterior membership 
probabilities combine information from pretreatment covariates and 
longitudinal outcomes; for future patients, the fitted gating model 
produces membership probabilities from pretreatment covariates alone.
The framework differs from existing methods in three respects. First, 
because the gating model enters the observed-data likelihood, 
pretreatment covariates contribute directly to subgroup discovery rather 
than being used only to predict subgroup labels estimated in a separate 
stage. Second, separating global component weights from 
covariate-dependent gating coefficients enables subgroup-number 
selection and patient-specific membership modeling within the same 
objective. A redundant component can therefore be eliminated globally 
without forcing membership probabilities among the retained groups to 
be constant across patients. Third, the longitudinal model allows both 
treatment-effect trajectories and covariance functions to vary across 
subgroups without prespecifying a parametric correlation structure. We 
establish selection consistency for the subgroup number, convergence 
rates for the functional estimators, and asymptotic normality of the 
finite-dimensional estimators.

The remainder of the paper is organized as follows. 
Section~\ref{sec:model} develops the proposed model and estimation 
procedure, including tuning-parameter selection and prospective 
membership prediction. Section~\ref{sec:theory} establishes 
subgroup-selection consistency and the asymptotic properties of the 
estimators. Section~\ref{sec:simulation} evaluates finite-sample 
performance through simulations. Section~\ref{sec:application} 
illustrates the method using data from the COMPETE migraine trial and 
the SANAD epilepsy study. The final section discusses the implications, 
limitations, and directions for future research.

\section{Model and estimation}\label{sec:model}

Consider $n$ independent subjects, with subject $i$ observed at times 
$t_{ij}\in[0,1]$, $j=1,\ldots,n_i$. Let $Y_i(t_{ij})$ and 
$\bX_{ij}\in\mathbb R^d$ denote the corresponding longitudinal response 
and covariate vector. Define \(
\bY_i=(Y_i(t_{i1}),\ldots,Y_i(t_{in_i}))^{\mathrm T},
\bX_i=(\bX_{i1},\ldots,\bX_{in_i})^{\mathrm T},
\bt_i=(t_{i1},\ldots,t_{in_i})^{\mathrm T}.
\)
Let $D_i\in\{0,1\}$ denote treatment assignment, with $D_i=1$ indicating 
the treatment of interest and $D_i=0$ the comparator. Let 
$\bZ_i\in\mathbb R^q$ contain pretreatment covariates used to predict 
subgroup membership for new subjects. The vectors $\bX_{ij}$ and 
$\bZ_i$ may share baseline covariates. The observed data for subject $i$ are
$\mathcal O_i=(\bY_i,\bX_i,\bt_i,D_i,\bZ_i)$.

Let $G_i\in\{1,\ldots,G_0\}$ denote the latent subgroup membership of 
subject $i$, where $G_0$ is the true but unknown number of subgroups. 
Conditional on $G_i=g$, we assume the subgroup-specific 
varying-coefficient model
\begin{equation}\label{eq:longitudinal_model}
Y_i(t_{ij})
=
\bX_{ij}^{\mathrm T}
\{\bbeta_{g1}(t_{ij})+D_i\bbeta_{g2}(t_{ij})\}
+e_{ig}(t_{ij})+\epsilon_{igj},
\end{equation}
where $\bbeta_{g1}(t)=(\beta_{g1,1}(t),\ldots,\beta_{g1,d}(t))^{\mathrm T}$
represents the time-varying covariate effects, and
$\bbeta_{g2}(t)=(\beta_{g2,1}(t),\ldots,\beta_{g2,d}(t))^{\mathrm T}$
represents the time-varying treatment effects in subgroup $g$.
For a fixed covariate profile $\bm x$, the corresponding treatment
contrast is $\tau_g(t\mid\bm x)=\bm x^{\mathrm T}\bbeta_{g2}(t).$

Within-subject dependence in subgroup $g$ is represented by a mean-zero 
Gaussian process $e_{ig}(t)$ with covariance function
$\Sigma_g(s,t)=\operatorname{Cov}\{e_{ig}(s),e_{ig}(t)\}$. Its 
Karhunen--Lo\`eve expansion is
$e_{ig}(t)=\sum_{k=1}^{\infty}V_{igk}\phi_{gk}(t)$
\citep{ash2014topics,james2000principal}, where $\phi_{gk}(t)$ is the 
$k$th eigenfunction of $\Sigma_g$. The eigenfunctions are orthonormal, 
satisfying
$\int_0^1\phi_{gk}(t)\phi_{g\ell}(t)\,dt=\I(k=\ell)$, and the scores are 
independent across $k$, with $V_{igk}\sim N(0,\lambda_{gk})$. We order 
the eigenvalues so that
$\lambda_{g1}\geq\lambda_{g2}\geq\cdots>0$ and assume
$\sum_{k=1}^{\infty}\lambda_{gk}<\infty$. In practice, we retain the 
first $K$ terms and define
$\bm V_{ig}=(V_{ig1},\ldots,V_{igK})^{\mathrm T}$ and
$\bm\phi_g(t)=(\phi_{g1}(t),\ldots,\phi_{gK}(t))^{\mathrm T}$. Let
$\bLambda_g=\operatorname{diag}(\lambda_{g1},\ldots,\lambda_{gK})$; then
$e_{ig}(t)\approx\bm V_{ig}^{\mathrm T}\bm\phi_g(t)$. The measurement 
errors satisfy
$\epsilon_{igj}\overset{\mathrm{ind}}{\sim}N(0,\sigma_g^2)$ and are 
independent of $e_{ig}$. Let $\bPhi_g(\bt_i)\in\mathbb R^{n_i\times K}$ denote the matrix with 
$j$th row $\bm\phi_g(t_{ij})^{\mathrm T}$. Under the rank-$K$ 
approximation, the covariance matrix of $\bY_i$ conditional on $G_i=g$ 
is $ \bSigma_{ig} = \bPhi_g(\bt_i)\bLambda_g\bPhi_g(\bt_i)^{\mathrm T} +\sigma_g^2\bI_{n_i}. $

Because \(G_0\) is unknown, we embed the true model in an
over-specified working model with \(\Gmax\geq G_0\) candidate subgroups.
Under this working model, \(G_i\in\{1,\ldots,\Gmax\}\), and we relate
subgroup membership to pretreatment covariates through
\begin{equation}\label{eq:gating}
\pi_{ig}:=\pi_g(\bZ_i)=\Prb(G_i=g\mid\bZ_i)
=
\frac{\omega_g\exp(\bZ_i^{\mathrm T}\boldeta_g)}{\sum_{h=1}^{\Gmax}\omega_h\exp(\bZ_i^{\mathrm T}\boldeta_h)},
\qquad g=1,\ldots,\Gmax,
\end{equation}
where \(\omega_g\geq0\), \(\sum_{g=1}^{\Gmax}\omega_g=1\), and
\(\boldeta_g\in\mathbb R^q\). The global weights determine which components remain active,
whereas the gating coefficients allow membership probabilities among
active components to vary with pretreatment covariates. In particular,
\(\omega_g=0\) removes subgroup \(g\) for every covariate profile,
whereas variation in \(\pi_{ig}\) represents subject-specific
membership rather than global subgroup removal.  Let \(\calA_0=\{g:\omega_{g0}>0\}\) denote the true active set, so that
\(|\calA_0|=G_0\). For identifiability, we exclude an intercept from
\(\bZ_i\) and set \(\boldeta_1=\bzero\). The global weights satisfy
\(\pi_g(\bzero)=\omega_g\). If \(\boldeta_g=\bzero\) for every \(g\),
the gating model reduces to constant mixing proportions, with
\(\pi_{ig}=\omega_g\) for all \(i\); in general, however, \(\omega_g\)
need not equal the marginal prevalence of subgroup \(g\).

\subsection{Penalized estimation}\label{sec:estimation}

Let $\bm\mu_{ig}=(\mu_{ig1},\ldots,\mu_{ign_i})^{\mathrm T}$ denote the
subgroup-specific mean vector, where
$\mu_{igj}=\bX_{ij}^{\mathrm T}
\{\bbeta_{g1}(t_{ij})+D_i\bbeta_{g2}(t_{ij})\}$.
Under the rank-$K$ covariance approximation, the subgroup-conditional
density is
$f_{ig}:=f_g(\bY_i\mid\bX_i,\bt_i,D_i)
=\varphi_{n_i}(\bY_i;\bm\mu_{ig},\bSigma_{ig})$,
where $\varphi_{n_i}$ denotes the $n_i$-variate Gaussian density.
Marginalizing over $G_i$ gives the observed-data density
$f_i(\bY_i\mid\bX_i,\bt_i,D_i,\bZ_i)
=\sum_{g=1}^{\Gmax}\pi_{ig}f_{ig}$.

We approximate the nonparametric functions
$\{\bbeta_{g1},\bbeta_{g2},\bm\phi_g:g=1,\ldots,\Gmax\}$
using B-spline basis expansions \citep{schumaker2007spline}. Define the
H\"older class
\begin{equation}\label{eq:function_class_F}
\mathcal F=\left\{f:[0,1]\to\mathbb R
|f^{(q_1)}(s)-f^{(q_1)}(t)|\leq c_0|s-t|^{q_2},
\quad s,t\in[0,1]\right\},
\end{equation}
where $q_1$ is a nonnegative integer, $q_2\in(0,1]$,
$r=q_1+q_2\geq2$, and $c_0>0$. Assume that
$\bbeta_{g1},\bbeta_{g2}\in\mathcal F^d$ and
$\bm\phi_g\in\mathcal F^K$. Let
$\bm b_n(t)=(b_1(t),\ldots,b_{q_n}(t))^{\mathrm T}$ be a
$q_n$-dimensional B-spline basis on $[0,1]$. We use
$\bbeta_{gm}(t)\approx\btheta_{ngm}^{\mathrm T}\bm b_n(t)$
for $m=1,2$ and
$\bm\phi_g(t)\approx\bgamma_{ng}^{\mathrm T}\bm b_n(t)$,
where $\btheta_{ngm}\in\mathbb R^{q_n\times d}$ and
$\bgamma_{ng}\in\mathbb R^{q_n\times K}$. Let
$\bm R_n=\int_0^1\bm b_n(t)\bm b_n(t)^{\mathrm T}\,dt$.
The constraints
$\bgamma_{ng}^{\mathrm T}\bm R_n\bgamma_{ng}=\bI_K$
ensure orthonormality. For an orthonormalized spline basis,
$\bm R_n=\bI_{q_n}$. 

Let $\bB_i$ be the $n_i\times q_n$ matrix whose $j$th row is
$\bm b_n(t_{ij})^{\mathrm T}$. Set
$\bu_{ij}=\bX_{ij}\otimes\bm b_n(t_{ij})$,
$\bU_i=(\bu_{i1},\ldots,\bu_{in_i})^{\mathrm T}$, and
$\bW_i=(\bU_i,D_i\bU_i)$. Let $\btheta_{ng}$ denote the vector
formed by stacking $\operatorname{vec}(\btheta_{ng1})$ and
$\operatorname{vec}(\btheta_{ng2})$. The approximated mean and
covariance are then $\bm\mu_{ig}=\bW_i\btheta_{ng},$
and
$\bSigma_{ig} =\bB_i\bgamma_{ng}\bLambda_g \bgamma_{ng}^{\mathrm T}\bB_i^{\mathrm T} +\sigma_g^2\bI_{n_i}.$
Substituting these approximations into the mixture density, the
observed-data log-likelihood is
\begin{equation*}
\ell_n(\bTheta)
=\sum_{i=1}^n\log\left\{
\sum_{g=1}^{\Gmax}\pi_{ig}f_{ig}
\right\},
\end{equation*}
where $\bTheta$ collects the spline coefficients, eigenvalues,
measurement-error variances, global weights, and gating coefficients.

To remove redundant candidate subgroups, we apply an adapted
logarithmic penalty to the global weights \citep{huang2017},
$$\mathcal P_{\lambda_n,\epsilon_n}(\bomega)
=n\lambda_n\sum_{g=1}^{\Gmax}
\log\{(\epsilon_n+\omega_g)/\epsilon_n\},$$
where $\lambda_n>0$ controls the degree of subgroup selection and
$\epsilon_n>0$ ensures that the penalty remains finite at zero.
The penalized log-likelihood is
\begin{equation}\label{eq:penloglik}
Q_n(\bTheta)=\ell_n(\bTheta)
-\mathcal P_{\lambda_n,\epsilon_n}(\bomega).
\end{equation}
The estimator maximizes \eqref{eq:penloglik} subject to the
weight-simplex constraint, the reference restriction
$\boldeta_1=\bzero$, and the covariance and orthonormality constraints.
The estimated number of subgroups is
$\widehat G=\#\{g:\widehat\omega_g>0\}$.

\subsection{Estimation algorithm}\label{sec:em}

We estimate \(\bTheta\) using a blockwise ascent procedure that combines
EM-type posterior weighting with direct optimization of the penalized
observed-data objective \(Q_n\). Each iteration first computes the
posterior subgroup probabilities and then updates the longitudinal and
gating parameters in separate blocks. The longitudinal block updates
the mean and covariance parameters, whereas the gating block jointly
updates the global weights and covariate effects. Candidate updates are
accepted only when they do not decrease \(Q_n\) beyond a numerical
tolerance. This structure accommodates the covariance and weight-simplex
constraints while retaining a common objective for all model components.

Let $\delta_{ig}=\I(G_i=g)$ denote the latent subgroup indicator.
Treating $\{\delta_{ig}:i=1,\ldots,n,\ g=1,\ldots,\Gmax\}$ as missing
data, the complete-data log-likelihood is
$\ell_n^c(\bTheta)=\sum_{i=1}^n\sum_{g=1}^{\Gmax}
\delta_{ig}\{\log f_{ig}+\log\pi_{ig}\}$.
Given the current estimate $\bTheta^{(t)}$, the E-step computes the
posterior responsibilities
\begin{equation*}
h_{ig}^{(t)}
=
\Prb(G_i=g\mid\mathcal O_i;\bTheta^{(t)})
=
\frac{\pi_{ig}^{(t)}f_{ig}^{(t)}}{\displaystyle\sum_{\ell=1}^{\Gmax}
\pi_{i\ell}^{(t)}f_{i\ell}^{(t)}}.
\end{equation*}
These probabilities combine pretreatment information through
$\pi_{ig}^{(t)}$ with longitudinal evidence through $f_{ig}^{(t)}$.
They serve as subject- and subgroup-specific weights in the subsequent
longitudinal updates. For notational simplicity, iteration superscripts
are suppressed below, and unspecified quantities on the right-hand
sides are evaluated at the current iterate.

For fixed responsibilities and covariance matrices, the mean-function
coefficients are updated by posterior-weighted generalized least
squares,
\begin{equation}\label{eq:theta-update}
\widetilde{\btheta}_{ng}
=
\left\{\sum_{i=1}^n h_{ig}
\bW_i^{\mathrm T}\bSigma_{ig}^{-1}\bW_i\right\}^{-1}
\sum_{i=1}^n h_{ig}
\bW_i^{\mathrm T}\bSigma_{ig}^{-1}\bY_i.
\end{equation}

We next update the covariance structure using the GLS residuals
$\widetilde{\br}_{ig}
=\bY_i-\bW_i\widetilde{\btheta}_{ng}$.
Let $\bgamma_{ngk}$ denote the $k$th column of $\bgamma_{ng}$.
With the remaining covariance parameters fixed, the covariance score
equation gives the unconstrained proposal
\begin{equation}\label{eq:gamma-update}
\widetilde{\bgamma}_{ngk}
=
\left\{\sum_{i=1}^n h_{ig}
\bB_i^{\mathrm T}\bSigma_{ig}^{-1}\bB_i\right\}^{-1}
\sum_{i=1}^n h_{ig}
\bB_i^{\mathrm T}\bSigma_{ig}^{-1}
\widetilde{\br}_{ig}\widetilde{\br}_{ig}^{\mathrm T}
\bSigma_{ig}^{-1}\bB_i\bgamma_{ngk},
\qquad k=1,\ldots,K.
\end{equation}

For the eigenvalue update, let
$\bm\phi_{igk}=\bB_i\bgamma_{ngk}$.
With the remaining covariance parameters fixed, a Fisher-scoring
proposal for $\lambda_{gk}$ is
\begin{equation}\label{eq:lambda-update}
\widetilde{\lambda}_{gk}
=
\lambda_{gk}
+
\frac{
\displaystyle\sum_{i=1}^n h_{ig}
\left\{
\bm\phi_{igk}^{\mathrm T}\bSigma_{ig}^{-1}
\widetilde{\br}_{ig}\widetilde{\br}_{ig}^{\mathrm T}
\bSigma_{ig}^{-1}\bm\phi_{igk}
-
\bm\phi_{igk}^{\mathrm T}\bSigma_{ig}^{-1}
\bm\phi_{igk}
\right\}
}{
\displaystyle\sum_{i=1}^n h_{ig}
\left\{
\bm\phi_{igk}^{\mathrm T}\bSigma_{ig}^{-1}
\bm\phi_{igk}
\right\}^{2}
},
\qquad k=1,\ldots,K.
\end{equation}

Similarly, with the functional covariance parameters fixed, a
Fisher-scoring proposal for the measurement-error variance is
\begin{equation}\label{eq:sigma-update}
\widetilde{\sigma}_g^2
=
\sigma_g^2
+
\frac{
\displaystyle\sum_{i=1}^n h_{ig}
\left\{
\widetilde{\br}_{ig}^{\mathrm T}
\bSigma_{ig}^{-2}\widetilde{\br}_{ig}
-
\operatorname{tr}(\bSigma_{ig}^{-1})
\right\}
}{
\displaystyle\sum_{i=1}^n h_{ig}
\operatorname{tr}(\bSigma_{ig}^{-2})
},
\qquad g=1,\ldots,\Gmax.
\end{equation}

The proposed eigenfunction coefficients are orthonormalized by QR
decomposition in the orthonormal spline basis. The resulting
eigenvalue--eigenfunction pairs are ordered by decreasing eigenvalue,
and each eigenfunction is given a consistent sign across iterations. We then
apply backtracking between the current and proposed covariance parameters,
reimposing these constraints at each trial step. A trial step is accepted
only if it does not decrease \(Q_n\) beyond the numerical tolerance;
otherwise, the current covariance parameters are retained.

Holding the subgroup-conditional densities \(f_{ig}\) fixed, we update
the global weights and gating coefficients by increasing
\begin{equation}\label{eq:gatingobjective}
Q_G(\bomega,\boldeta)
=
\sum_{i=1}^n
\log\left\{
\sum_{g=1}^{\Gmax}\pi_{ig}f_{ig}
\right\}
-
\mathcal P_{\lambda_n,\epsilon_n}(\bomega),
\end{equation}
subject to the weight-simplex constraint and
\(\boldeta_1=\bzero\). This is the restriction of \(Q_n\) to the
gating parameters with the longitudinal parameters held fixed.

For the gating updates, define
\(\rho_{ig}=\pi_{ig}f_{ig}/
\{\sum_{\ell=1}^{\Gmax}\pi_{i\ell}f_{i\ell}\}\).
Unlike the fixed E-step responsibility \(h_{ig}\), \(\rho_{ig}\)
varies with the candidate gating parameters. The relevant derivatives
are
\begin{equation}\label{eq:gating-gradients}
\begin{aligned}
\frac{\partial Q_G}{\partial\boldeta_g}
&=
\sum_{i=1}^n
(\rho_{ig}-\pi_{ig})\bZ_i,
\qquad g\ \text{nonreference},\\
\frac{\partial Q_G}{\partial\omega_g}
&=
\frac{1}{\omega_g}
\sum_{i=1}^n(\rho_{ig}-\pi_{ig})
-
\frac{n\lambda_n}{\epsilon_n+\omega_g},
\qquad \omega_g>0.
\end{aligned}
\end{equation}

We update the global weights and nonreference gating coefficients by
constrained Newton-type optimization of \(Q_G\), subject to the
weight-simplex constraint, the reference restriction
\(\boldeta_1=\bzero\), and any specified coordinate bounds. A
backtracking line search is used to ensure that each accepted step does
not decrease \(Q_G\). After each update, components with weights at or
below the numerical threshold \(10^{-6}\) are removed, and the
remaining weights are renormalized. The active component with the
largest weight is then selected as the reference, and the gating
coefficients are recentered accordingly.

Algorithm~\ref{alg:joint-estimation} summarizes the procedure for fixed
\((q_n,K,\lambda_n)\).

\begin{algorithm}[!htbp]
\caption{Blockwise estimation for fixed \((q_n,K,\lambda_n)\)}
\label{alg:joint-estimation}
\small
\begin{algorithmic}[1]
\Require Data \(\{\mathcal O_i\}_{i=1}^n\), \(\Gmax\), tuning parameters, and initial values.
\Ensure \(\widehat\bTheta\) and the active set \(\widehat{\mathcal A}\).
\State Construct the orthonormal spline basis and design matrices.
\For{each initialization}
    \State Initialize model parameters and the active subgroup set.
    \Repeat
        \State Compute the posterior responsibilities \(h_{ig}\).
        \State Update the mean parameters using \eqref{eq:theta-update}.
        \State Update the covariance parameters using \eqref{eq:gamma-update}--\eqref{eq:sigma-update} and backtracking.
        \State Update the gating parameters by constrained optimization of \eqref{eq:gatingobjective}.
        \State Prune inactive components and update the active subgroup set.
        \State Compute \(\Delta_Q^{(t)}=\dfrac{|Q_n^{(t)}-Q_n^{(t-1)}|}{1+|Q_n^{(t-1)}|}\) and the blockwise first-order residual \(R^{(t)}\).
    \Until{\(\Delta_Q^{(t)}\leq\epsilon_Q\) and \(R^{(t)}\leq\epsilon_R\), or the iteration limit is reached}
    \State Store the fitted parameters and objective value.
\EndFor
\State Retain the fit with the largest attained value of \(Q_n\).
\State \Return \(\widehat\bTheta\) and \(\widehat{\mathcal A}\).
\end{algorithmic}
\end{algorithm}

\subsection{Tuning-parameter selection}\label{sec:tuning}

The tuning parameters are the spline dimension \(q_n\), covariance rank
\(K\), and penalty parameter \(\lambda_n\). For each candidate
configuration, we compute
\(\operatorname{BIC}=-2\ell_n(\widehat\bTheta)
+\operatorname{df}(\widehat\bTheta)\log n\), where \(\ell_n\) is the
unpenalized observed-data log-likelihood. For a fitted model with
\(\widehat G\) active subgroups, the nominal parameter count is
\begin{equation*}
\operatorname{df}(\widehat\bTheta)
=\widehat G\left\{
2dq_n+q_nK-\frac{K(K+1)}{2}+K+1
\right\}
+(\widehat G-1)(q+1).
\end{equation*}
The terms correspond to the varying-coefficient functions, orthonormal
eigenfunctions, eigenvalues, measurement-error variances, global
weights, and nonreference gating coefficients. We select the
configuration with the smallest BIC.

\subsection{Subgroup membership estimation and prediction}\label{sec:prediction}

For a subject in the training sample, subgroup membership is inferred
from
\[
\Prb(G_i=g\mid\mathcal O_i;\widehat\bTheta)
=
\frac{\widehat\pi_g(\bZ_i)\widehat f_{ig}}{\sum_{h=1}^{\widehat G}\widehat\pi_h(\bZ_i)\widehat f_{ih}}.
\]
These posterior probabilities incorporate both pretreatment covariates
and the observed longitudinal response. For a new subject with
pretreatment covariates \(\bm z_*\), only pretreatment information is
available. The fitted gating model then gives
\begin{equation}\label{eq:newpi}
\widehat\pi_g(\bm z_*)
=
\frac{\widehat\omega_g
\exp(\bm z_*^{\mathrm T}\widehat{\boldeta}_g)}{\displaystyle\sum_{h\in\widehat{\mathcal A}}
\widehat\omega_h
\exp(\bm z_*^{\mathrm T}\widehat{\boldeta}_h)},
\qquad g\in\widehat{\mathcal A},
\end{equation}
where \(\widehat{\mathcal A}=\{g:\widehat\omega_g>0\}\). The subgroup with the largest probability gives a prediction,
whereas the full probability vector quantifies assignment uncertainty.

\section{Asymptotic properties}\label{sec:theory}

Let
\(
\widehat{\bm\Omega}_n
=
\left(
\widehat\bomega,\widehat{\boldeta},
\left\{
\widehat{\bbeta}_{g1},\widehat{\bbeta}_{g2},
\widehat{\bm\phi}_g,\widehat{\bLambda}_g,
\widehat{\sigma}_g^2
\right\}_{g=1}^{\widehat G}
\right),
\)
where
\(\widehat{\bbeta}_{gm}(t)
=\widehat{\btheta}_{ngm}^{\mathrm T}\bm b_n(t)\), \(m=1,2\), and
\(\widehat{\bm\phi}_g(t)
=\widehat{\bgamma}_{ng}^{\mathrm T}\bm b_n(t)\). The estimated covariance
function is
\[
\widehat\Sigma_g(s,t)
=
\sum_{k=1}^{K}
\widehat\lambda_{gk}\widehat\phi_{gk}(s)\widehat\phi_{gk}(t)
+\widehat\sigma_g^2\I(s=t).
\]
Let \(\|\cdot\|\) and \(\|\cdot\|_2\) denote the Euclidean and \(L^2\)
norms, respectively. The working parameter space is
\[
\bm\Omega^*
=
[0,1]^{\Gmax}
\times\mathbb R^{q(\Gmax-1)}
\times\mathcal F^{2d\Gmax}
\times\mathcal F^{K\Gmax}
\times\mathbb R_+^{K\Gmax}
\times\mathbb R_+^{\Gmax},
\]
with the weight-simplex, reference, and orthonormality constraints
specified in Section~\ref{sec:model}, where
\(\mathbb R_+=(0,\infty)\).

For \(\bm\Gamma=(\bomega,\bbeta,\bm\Sigma)\), define the distance between
two \(G_0\)-component parameter values, after aligning their component
labels, by
\[
d(\bm\Gamma_1,\bm\Gamma_2)
=
\left\{
\|\bomega_1-\bomega_2\|^2
+\sum_{g=1}^{G_0}\sum_{m=1}^{2}
\|\bbeta_{gm,1}-\bbeta_{gm,2}\|_2^2
+\sum_{g=1}^{G_0}
\|\Sigma_{g,1}-\Sigma_{g,2}\|_2^2
\right\}^{1/2}.
\]
A subscript \(0\) denotes a true parameter or function. Without loss of
generality, order the global weights so that
\(\omega_{1,0}\geq\cdots\geq\omega_{G_0,0}>0\) and
\(\omega_{G_0+1,0}=\cdots=\omega_{\Gmax,0}=0\). The following conditions
are used to establish the asymptotic properties of
\(\widehat{\bm\Omega}_n\).

\begin{itemize}[leftmargin=3em]
    \item[(A1)] Let
    \(0=t_0<t_1<\cdots<t_{M_n}<t_{M_n+1}=1\) denote the spline-knot
    sequence. Then
    \(\max_{1\leq j\leq M_n+1}(t_j-t_{j-1})=O(n^{-\kappa})\) for
    \(0<\kappa<1/2\), and the ratio of the largest to the smallest knot
    spacing is bounded.
    \item[(A2)] The parameters associated with the \(G_0\) active
    components lie in the interior of their corresponding parameter
    spaces.
    \item[(A3)] \(K=O(n^e)\), where
    \(0\leq e<\min(1-\kappa,2r\kappa)\).
    \item[(A4)] For each active subgroup,
    \(\sum_{k=1}^{\infty}\lambda_{gk}<\infty\).
    \item[(A5)] The matrix
    \(\E\{\bm S(\bzeta_0)\bm S(\bzeta_0)^{\mathrm T}\}\) is finite and
    positive definite, where
    \(\bzeta=(\bm\sigma^2,\bomega_{-1},\boldeta_{2:G_0})^{\mathrm T}\).
    Here \(\bomega_{-1}=(\omega_2,\ldots,\omega_{G_0})^{\mathrm T}\),
    \(\boldeta_{2:G_0}\) stacks the nonreference gating coefficients,
    and \(\boldeta_1=\bzero\). The score \(\bm S(\bzeta_0)\) is defined
    in the proof of Theorem~\ref{thm:normality}.
\end{itemize}

Condition (A1) excludes highly unbalanced knot sequences and is standard
in spline analysis \citep{lu2009semiparametric}. Condition (A2) excludes
degenerate active-component parameters \citep{chen2010varying}.
Condition (A3) controls the growth of the covariance rank, which is
typically small in practice. Condition (A4) ensures that each covariance
operator is trace class \citep{hall2007methodology}. Condition (A5)
ensures that the asymptotic covariance matrix is well defined
\citep{ma2015sieve}.

\begin{theorem}\label{thm:selection}
Suppose Conditions (A1)--(A5) hold,
\(\lambda_n\sqrt n\to0\),
\(\lambda_n\sqrt n\log n\to\infty\), and
\(\epsilon_n=o\{1/(\sqrt n\log n)\}\). Then
\(\Prb(\widehat G=G_0)\to1\) as \(n\to\infty\).
\end{theorem}

\begin{theorem}\label{thm:estimation}
Suppose Conditions (A1)--(A5) hold,
\(\lambda_n\sqrt n\to0\), and
\(\epsilon_n=o\{1/(\sqrt n\log n)\}\). Then
\[
d(\widehat{\bm\Gamma}_n,\bm\Gamma_0)
=
O_p\left[
n^{-\min\{(1-\kappa-e)/2,\;r\kappa-e/2\}}
\right].
\]
Here \(r\) is the smoothness index defined in
\eqref{eq:function_class_F}, and the knot sequence grows at rate
\(n^\kappa\). Increasing \(e\) enlarges the covariance parameter space
and slows the convergence rate. If \(e=0\), so that \(K\) is fixed, then
\[
d(\widehat{\bm\Gamma}_n,\bm\Gamma_0)
=
O_p\left[
n^{-\min\{(1-\kappa)/2,\;r\kappa\}}
\right].
\]
In particular, choosing \(\kappa=1/(2r+1)\) yields
\(d(\widehat{\bm\Gamma}_n,\bm\Gamma_0)
=O_p\{n^{-r/(2r+1)}\}\), the optimal rate for univariate nonparametric
function estimation.
\end{theorem}

\begin{theorem}\label{thm:normality}
Suppose Conditions (A1)--(A5) hold with \(r\geq2\),
\(1/(4r)<\kappa<1/2\), \(\lambda_n\sqrt n\to0\), and
\(\epsilon_n=o\{1/(\sqrt n\log n)\}\). For
\(\bzeta=(\bm\sigma^2,\bomega_{-1},\boldeta_{2:G_0})^{\mathrm T}\),
\[
\sqrt n(\widehat{\bzeta}-\bzeta_0)
\ \xrightarrow{d}\
N\{0,\bm I^{-1}(\bzeta_0)\},
\]
where
\(\bm I(\bzeta_0)
=\E\{\bm S(\bzeta_0)\bm S(\bzeta_0)^{\mathrm T}\}\).
\end{theorem}

In particular,
\[
\sqrt n(\widehat\bomega_{-1}-\bomega_{-1,0})
\ \xrightarrow{d}\
N\left\{0,[\bm I^{-1}(\bzeta_0)]_{\omega\omega}\right\}.
\]
Here \([\bm I^{-1}(\bzeta_0)]_{\omega\omega}\) is the block corresponding
to \(\bomega_{-1}\). Because
\(\omega_1=1-\sum_{g=2}^{G_0}\omega_g\), this result also determines the
asymptotic distribution of the full vector of active global weights.

\section{Simulation study}\label{sec:simulation}

We evaluate the finite-sample performance of the proposed joint method
through Monte Carlo simulations. The primary comparator is a two-step
procedure that first identifies longitudinal subgroups and then relates
estimated membership to pretreatment covariates \citep{zhang2026}. We
also include an oracle estimator based on the true subgroup labels and
a latent class mixed-model benchmark (HLME) \citep{lin2000}. We assess
subgroup-number selection, longitudinal parameter estimation, training
membership recovery, and pretreatment membership prediction for new
subjects.

\subsection{Simulation design}

Each replication includes a training sample with
\(n_{\mathrm{train}}\in\{600,800\}\) and an independent test sample with
\(n_{\mathrm{test}}=400\). The true subgroup number is \(G_0=2\), and
the candidate upper bound is \(\Gmax\in\{5,8,10\}\). We use \(R=200\)
independent replications for each setting. Treatment is generated as
\(D_i\sim\operatorname{Bernoulli}(0.5)\), and
\(n_i\sim\operatorname{DiscreteUniform}\{8,9,10,11,12\}\). Conditional
on \(n_i\), observation times are sampled independently from
\(\operatorname{Uniform}(0,1)\) and then ordered within subject.

The longitudinal covariate vector is
\(\bm X_{ij}=\{X_{i1}(t_{ij}),X_{i2}(t_{ij}),
X_{i3}(t_{ij})\}^{\mathsf T}\), where \(X_{i1}(t)=1\),
\(X_{i2}(t)=A_i\), and \(A_i\sim N(1,1)\). Its third component is
\[
X_{i3}(t)=\frac{1}{2}+\sin(2\pi t)+\sum_{m=1}^{10}\xi_{im}\psi_m(t),
\qquad \psi_m(t)=\sqrt2\cos(m\pi t),\qquad
\xi_{im}\sim\mathcal N(0,4m^{-2}).
\]
The scalar \(A_i\) is generated once per subject. The coefficients
\(\xi_{im}\) are independent across subjects and basis indices. The
pretreatment predictors are generated independently as
\(Z_{i1},Z_{i2}\sim N(1,4)\). Define
\(\widetilde Z_{i1}=Z_{i1}-1\),
\(\widetilde Z_{i2}=Z_{i2}-1\), and
\(\widetilde{\bm Z}_i
=(\widetilde Z_{i1},\widetilde Z_{i2})^{\mathsf T}\).
The latent subgroup label is generated according to
\begin{equation}\label{eq:simgating}
p_i:=\Pr(G_i=1\mid\bm Z_i)
=\frac{\exp\{c(\widetilde Z_{i1}+\widetilde Z_{i2})\}}{1+\exp\{c(\widetilde Z_{i1}+\widetilde Z_{i2})\}},
\qquad \Pr(G_i=2\mid\bm Z_i)=1-p_i,
\end{equation}
with \(c=2\). Because the linear predictor is symmetric about zero,
both subgroups have marginal prevalence \(0.5\). Under the proposed
parameterization, the true gating parameters are
\(\omega_1=\omega_2=0.5\), \(\bm\eta_1=\bm0\), and
\(\bm\eta_2=(-2,-2)^{\mathsf T}\).

Conditional on $G_i=g$, the longitudinal response follows
\begin{equation*}
Y_i(t_{ij})
=\bm X_{ij}^{\mathsf T}\{\bm\beta_{g1}(t_{ij})+D_i\bm\beta_{g2}(t_{ij})\}
+\bm V_{ig}^{\mathsf T}\bm\phi_g(t_{ij})+\epsilon_{igj},
\end{equation*}
where \(\bm V_{ig}\sim N_2(\bm0,\bm\Lambda_g)\) and
\(\epsilon_{igj}\sim N(0,\sigma_g^2)\) are mutually independent. We set
\(\bm\Lambda_1=\operatorname{diag}(1,0.25)\),
\(\bm\Lambda_2=\operatorname{diag}(1.2,0.30)\), and
\(\sigma_1^2=\sigma_2^2=0.25\). Let
\(\bm1_3=(1,1,1)^{\mathsf T}\). The subgroup-specific coefficient
functions are
\begin{align*}
\bm\beta_{11}(t)&=(1+0.45t)\bm1_3,
&\bm\beta_{12}(t)&=(0.4-0.35\exp(t))\bm1_3,\\
\bm\beta_{21}(t)&=(1+0.15t)\bm1_3,
&\bm\beta_{22}(t)&=(0.4-0.10\exp(t))\bm1_3,
\end{align*}
and
\(\bm\phi_1(t)
=\{\sqrt2\cos(\pi t),\sqrt2\sin(\pi t)\}^{\mathsf T}\), with
\(\bm\phi_2(t)=\bm\phi_1(t)\).

Both penalized procedures use the same orthonormalized cubic B-spline
basis, with interior knots at \(0.25\), \(0.50\), and \(0.75\)
(\(q_n=7\)), and covariance rank \(K=2\). We use \(\lambda=0.02\) for
the joint estimator and \(\lambda=0.04\) for the first stage of the
two-step procedure. Joint-model gating coefficients are constrained to
\([-5,5]\), whereas the second-stage logistic coefficients are
unconstrained. Each longitudinal fit is allowed at most 200 iterations.

\subsection{Competing methods and implementation}

The joint estimator fits the
longitudinal and membership models simultaneously using
Algorithm~\ref{alg:joint-estimation}. The two-step comparator first fits
a longitudinal mixture with constant component weights. It then treats
the posterior membership probabilities as fixed soft responses in a
multinomial logistic regression on the centered pretreatment
predictors. The second stage does not update the longitudinal
parameters. For training classification, the fitted gating model is combined
with the fixed component densities. Predictions for new subjects use
the gating model alone. This implementation uses soft-label logistic regression
rather than the posterior-log-odds regression of \citet{zhang2026}.

The oracle estimator fits two groups using the true training labels.
HLME fits two latent classes using \texttt{hlme} from the R package
\texttt{lcmm} \citep{proustlima2017}. It uses the same spline mean
design and pretreatment membership predictors. Its random-effects model
uses the true cosine and sine functions, a diagonal covariance, a
class-specific proportional factor, and a common residual variance.
Thus, HLME is supplied with the true subgroup number and covariance
shapes; it neither estimates eigenfunctions nor selects the subgroup
number.

\subsection{Evaluation criteria}

We report the number of replications that correctly identify \(G_0=2\) out of 200 (RCI). Training membership recovery is measured by IIC, the proportion
correctly classified after label alignment. For new subjects, we report
classification accuracy (ACC), Brier score, negative log predictive
probability (NLP), and
\(\operatorname{RMSE}_p
=\{n_{\mathrm{test}}^{-1}\sum_i(\widehat p_i-p_i)^2\}^{1/2}\),
where \(p_i\) is defined in \eqref{eq:simgating}. Higher ACC and lower
Brier score, NLP, and \(\operatorname{RMSE}_p\) indicate better
prediction. All test predictions use pretreatment covariates alone.
Parameter results are reported as RMSE (bias), with signed bias for
scalar parameters. For a function evaluated at \(M=200\) equally spaced
grid points, functional bias is
\(\{M^{-1}\sum_m[\overline{\widehat f}(t_m)-f_0(t_m)]^2\}^{1/2}\).
Functional RMSE averages squared errors over grid points and
replications. 


\subsection{Simulation results}
Tables~\ref{tab:simulation-final-n600} and
\ref{tab:simulation-final-n800} summarize the results for \(n=600\)
and \(n=800\). Each table compares the oracle estimator, the
fixed-two-group HLME benchmark, and both penalized procedures initialized
at \(\Gmax=5,8,10\). The comparison involved 3,200 fits across 400 paired training and test
datasets, and all fits completed without execution errors. All oracle
and HLME fits met their numerical convergence criteria. Returned
estimates from the penalized procedures that did not meet the numerical
convergence criterion are retained in the summaries.

For \(n=600\), the joint estimator selected two subgroups in 196, 194,
and 192 replications for \(\Gmax=5,8,10\), respectively. The
corresponding counts for the two-step procedure were 196, 180, and 191.
For \(n=800\), these counts were 200, 200, and 198 for the joint
estimator and 193, 194, and 195 for the two-step procedure. Thus, the
joint estimator had equal or higher correct-selection counts in every
configuration considered, indicating that the global-weight penalty
effectively removed redundant candidate subgroups across the three
choices of \(\Gmax\).

Across the candidate upper bounds, the joint estimator's errors in the
longitudinal parameters were stable. Its varying-coefficient RMSEs were
consistently smaller than those of the two-step procedure, and its
eigenfunction RMSEs were also generally smaller. This pattern is
consistent with more accurate recovery of the longitudinal component
when pretreatment covariates inform subgroup learning.

Among fits selecting two subgroups, the joint estimator achieved IIC
values of 95.42--95.44\% for \(n=600\) and approximately 95.61\% for
\(n=800\). The corresponding ranges for the two-step procedure were
67.17--69.36\% and 65.08--65.88\%. Test-set classification accuracy
ranged from 90.48--90.59\% for the joint estimator and
79.60--81.78\% for the two-step procedure. The corresponding Brier score
ranges were 0.068--0.069 and 0.182--0.194, and test-set
\(\operatorname{RMSE}_p\) ranged from 0.034 to 0.042 and from 0.331 to
0.350, respectively. Bias in the gating slopes also showed substantially
less attenuation under the joint estimator. Thus, the improvement was
evident in training-set membership recovery, test-set classification,
and probability estimation. These comparisons remain conditional on
selecting two subgroups and use method-specific eligible replications.

The correctly specified HLME benchmark had IIC values of 95.53\% and
95.70\% for \(n=600\) and \(n=800\), respectively. Its prediction
performance was close to that of the joint estimator. HLME often had
smaller covariance and gating-parameter RMSEs, reflecting the additional
subgroup-number and covariance-shape information supplied to this
benchmark. Accordingly, the joint estimator did not outperform HLME
for every parameter. The oracle estimator provided a known-membership
reference for longitudinal estimation. Conditional selection and Monte
Carlo variation can produce small departures from a strict empirical
RMSE ordering.

For \(n=600\) and \(\Gmax=5\), the average estimated eigenfunctions and
varying-coefficient functions followed their true shapes.
Figure~\ref{fig:simulation-n600-g5-eigen} displays the four
subgroup-specific eigenfunctions, and
Figure~\ref{fig:simulation-n600-g5-beta-combined} displays the six
coefficient functions for each subgroup. Solid curves denote the true
functions, and dashed curves denote averages across replications. The
pointwise bands summarize empirical variation across the 196
replications selecting two subgroups after label and eigenfunction
alignment. They are neither standard errors of the Monte Carlo averages
nor simultaneous confidence bands. Variation was greater near the
boundaries for several coefficient functions.

\begingroup
\makeatletter
\setlength{\@fptop}{0pt}
\makeatother
\clearpage
\begin{landscape}
\begin{table}[p]
\centering
\makebox[\linewidth][c]{%
\begin{minipage}{\linewidth}
\captionsetup{font=small,labelfont=bf,labelsep=simulationbar,justification=raggedright,singlelinecheck=false,name=TABLE}
\caption{Simulation results for $n=600$ (200 replications per method).}
\label{tab:simulation-final-n600}
\begingroup
\fontsize{8.5}{10.2}\selectfont
\setlength{\tabcolsep}{2.3pt}
\renewcommand{\arraystretch}{0.86}
\setlength{\heavyrulewidth}{0.65pt}
\setlength{\lightrulewidth}{0.35pt}
\setlength{\cmidrulewidth}{0.3pt}
\resizebox{\linewidth}{!}{%
\begin{tabular}{@{}lrr@{\hspace{8pt}}rr@{\hspace{8pt}}rr@{\hspace{8pt}}rr@{}}
\toprule
\multirow{2}{*}{Parameter} & \multicolumn{1}{c}{\multirow{2}{*}{\textbf{Oracle}}} & \multicolumn{1}{c}{\multirow{2}{*}{\textbf{HLME} $(G=2)$}} & \multicolumn{2}{c}{\boldmath$G=5$} & \multicolumn{2}{c}{\boldmath$G=8$} & \multicolumn{2}{c}{\boldmath$G=10$} \\
\cmidrule(lr){4-5}\cmidrule(lr){6-7}\cmidrule(l){8-9}
 & & & \multicolumn{1}{c}{\textbf{Proposed}} & \multicolumn{1}{c}{\textbf{Two-step}} & \multicolumn{1}{c}{\textbf{Proposed}} & \multicolumn{1}{c}{\textbf{Two-step}} & \multicolumn{1}{c}{\textbf{Proposed}} & \multicolumn{1}{c}{\textbf{Two-step}} \\
\midrule
$\lambda_{11}$ & $0.080\,(-0.012)$ & $0.072\,(-0.015)$ & $0.083\,(-0.010)$ & $0.136\,(\phantom{-}0.043)$ & $0.082\,(-0.011)$ & $0.129\,(\phantom{-}0.034)$ & $0.082\,(-0.009)$ & $0.137\,(\phantom{-}0.041)$ \\
$\lambda_{12}$ & $0.023\,(-0.008)$ & $0.021\,(-0.008)$ & $0.025\,(-0.009)$ & $0.031\,(\phantom{-}0.002)$ & $0.025\,(-0.008)$ & $0.031\,(\phantom{-}0.000)$ & $0.025\,(-0.008)$ & $0.029\,(\phantom{-}0.000)$ \\
$\lambda_{21}$ & $0.118\,(-0.008)$ & $0.102\,(-0.010)$ & $0.120\,(-0.010)$ & $0.157\,(-0.074)$ & $0.121\,(-0.010)$ & $0.149\,(-0.064)$ & $0.120\,(-0.008)$ & $0.157\,(-0.072)$ \\
$\lambda_{22}$ & $0.028\,(-0.007)$ & $0.024\,(-0.007)$ & $0.029\,(-0.007)$ & $0.042\,(-0.024)$ & $0.029\,(-0.007)$ & $0.042\,(-0.023)$ & $0.030\,(-0.008)$ & $0.041\,(-0.023)$ \\
\addlinespace[1.8pt]
$\omega_{1}$ & \multicolumn{1}{c}{/} & $0.057\,(\phantom{-}0.001)$ & $0.101\,(\phantom{-}0.004)$ & $0.040\,(\phantom{-}0.001)$ & $0.096\,(\phantom{-}0.007)$ & $0.040\,(-0.002)$ & $0.094\,(\phantom{-}0.005)$ & $0.038\,(\phantom{-}0.000)$ \\
$\omega_{2}$ & \multicolumn{1}{c}{/} & $0.057\,(-0.001)$ & $0.101\,(-0.004)$ & $0.040\,(-0.001)$ & $0.096\,(-0.007)$ & $0.040\,(\phantom{-}0.002)$ & $0.094\,(-0.005)$ & $0.038\,(\phantom{-}0.000)$ \\
$\eta_{21}$ & \multicolumn{1}{c}{/} & $0.381\,(-0.084)$ & $0.555\,(-0.148)$ & $1.841\,(\phantom{-}1.835)$ & $0.516\,(-0.131)$ & $1.819\,(\phantom{-}1.812)$ & $0.498\,(-0.137)$ & $1.826\,(\phantom{-}1.819)$ \\
$\eta_{22}$ & \multicolumn{1}{c}{/} & $0.376\,(-0.087)$ & $0.553\,(-0.153)$ & $1.838\,(\phantom{-}1.832)$ & $0.509\,(-0.138)$ & $1.814\,(\phantom{-}1.807)$ & $0.498\,(-0.145)$ & $1.820\,(\phantom{-}1.813)$ \\
$\sigma^2_{1}$ & $0.008\,(-0.004)$ & $0.006\,(-0.003)$ & $0.009\,(-0.005)$ & $0.010\,(-0.005)$ & $0.009\,(-0.005)$ & $0.010\,(-0.005)$ & $0.009\,(-0.005)$ & $0.009\,(-0.004)$ \\
$\sigma^2_{2}$ & $0.008\,(-0.004)$ & $0.006\,(-0.003)$ & $0.009\,(-0.004)$ & $0.011\,(-0.006)$ & $0.009\,(-0.004)$ & $0.011\,(-0.006)$ & $0.009\,(-0.004)$ & $0.011\,(-0.006)$ \\
\addlinespace[1.8pt]
$\phi_{11}(t)$ & $0.047\,(\phantom{-}0.002)$ & \multicolumn{1}{c}{/} & $0.051\,(\phantom{-}0.003)$ & $0.062\,(\phantom{-}0.003)$ & $0.050\,(\phantom{-}0.002)$ & $0.060\,(\phantom{-}0.003)$ & $0.051\,(\phantom{-}0.003)$ & $0.062\,(\phantom{-}0.003)$ \\
$\phi_{12}(t)$ & $0.065\,(\phantom{-}0.004)$ & \multicolumn{1}{c}{/} & $0.070\,(\phantom{-}0.004)$ & $0.080\,(\phantom{-}0.005)$ & $0.069\,(\phantom{-}0.004)$ & $0.079\,(\phantom{-}0.004)$ & $0.069\,(\phantom{-}0.005)$ & $0.079\,(\phantom{-}0.004)$ \\
$\phi_{21}(t)$ & $0.047\,(\phantom{-}0.005)$ & \multicolumn{1}{c}{/} & $0.049\,(\phantom{-}0.005)$ & $0.060\,(\phantom{-}0.004)$ & $0.054\,(\phantom{-}0.004)$ & $0.058\,(\phantom{-}0.004)$ & $0.049\,(\phantom{-}0.004)$ & $0.060\,(\phantom{-}0.004)$ \\
$\phi_{22}(t)$ & $0.062\,(\phantom{-}0.006)$ & \multicolumn{1}{c}{/} & $0.065\,(\phantom{-}0.007)$ & $0.081\,(\phantom{-}0.006)$ & $0.069\,(\phantom{-}0.005)$ & $0.078\,(\phantom{-}0.005)$ & $0.065\,(\phantom{-}0.006)$ & $0.080\,(\phantom{-}0.006)$ \\
\addlinespace[1.8pt]
$\beta_{111}(t)$ & $0.137\,(\phantom{-}0.008)$ & $0.148\,(\phantom{-}0.011)$ & $0.148\,(\phantom{-}0.012)$ & $0.254\,(\phantom{-}0.123)$ & $0.148\,(\phantom{-}0.012)$ & $0.257\,(\phantom{-}0.126)$ & $0.146\,(\phantom{-}0.010)$ & $0.253\,(\phantom{-}0.115)$ \\
$\beta_{112}(t)$ & $0.101\,(\phantom{-}0.005)$ & $0.106\,(\phantom{-}0.005)$ & $0.107\,(\phantom{-}0.005)$ & $0.202\,(\phantom{-}0.131)$ & $0.107\,(\phantom{-}0.006)$ & $0.194\,(\phantom{-}0.102)$ & $0.107\,(\phantom{-}0.005)$ & $0.194\,(\phantom{-}0.119)$ \\
$\beta_{113}(t)$ & $0.020\,(\phantom{-}0.001)$ & $0.021\,(\phantom{-}0.001)$ & $0.021\,(\phantom{-}0.001)$ & $0.148\,(\phantom{-}0.125)$ & $0.021\,(\phantom{-}0.001)$ & $0.141\,(\phantom{-}0.114)$ & $0.021\,(\phantom{-}0.001)$ & $0.143\,(\phantom{-}0.116)$ \\
$\beta_{121}(t)$ & $0.215\,(\phantom{-}0.015)$ & $0.222\,(\phantom{-}0.016)$ & $0.221\,(\phantom{-}0.020)$ & $0.308\,(\phantom{-}0.120)$ & $0.221\,(\phantom{-}0.017)$ & $0.320\,(\phantom{-}0.119)$ & $0.222\,(\phantom{-}0.019)$ & $0.311\,(\phantom{-}0.111)$ \\
$\beta_{122}(t)$ & $0.152\,(\phantom{-}0.008)$ & $0.156\,(\phantom{-}0.007)$ & $0.156\,(\phantom{-}0.008)$ & $0.238\,(\phantom{-}0.134)$ & $0.156\,(\phantom{-}0.008)$ & $0.240\,(\phantom{-}0.107)$ & $0.156\,(\phantom{-}0.009)$ & $0.236\,(\phantom{-}0.125)$ \\
$\beta_{123}(t)$ & $0.028\,(\phantom{-}0.001)$ & $0.030\,(\phantom{-}0.001)$ & $0.030\,(\phantom{-}0.002)$ & $0.149\,(\phantom{-}0.125)$ & $0.030\,(\phantom{-}0.001)$ & $0.142\,(\phantom{-}0.114)$ & $0.030\,(\phantom{-}0.001)$ & $0.146\,(\phantom{-}0.118)$ \\
$\beta_{211}(t)$ & $0.154\,(\phantom{-}0.012)$ & $0.162\,(\phantom{-}0.015)$ & $0.163\,(\phantom{-}0.016)$ & $0.268\,(\phantom{-}0.122)$ & $0.162\,(\phantom{-}0.017)$ & $0.256\,(\phantom{-}0.118)$ & $0.163\,(\phantom{-}0.013)$ & $0.255\,(\phantom{-}0.112)$ \\
$\beta_{212}(t)$ & $0.106\,(\phantom{-}0.007)$ & $0.110\,(\phantom{-}0.008)$ & $0.111\,(\phantom{-}0.008)$ & $0.212\,(\phantom{-}0.132)$ & $0.110\,(\phantom{-}0.008)$ & $0.196\,(\phantom{-}0.107)$ & $0.111\,(\phantom{-}0.006)$ & $0.208\,(\phantom{-}0.121)$ \\
$\beta_{213}(t)$ & $0.020\,(\phantom{-}0.001)$ & $0.020\,(\phantom{-}0.001)$ & $0.021\,(\phantom{-}0.001)$ & $0.148\,(\phantom{-}0.125)$ & $0.021\,(\phantom{-}0.001)$ & $0.141\,(\phantom{-}0.113)$ & $0.021\,(\phantom{-}0.001)$ & $0.142\,(\phantom{-}0.116)$ \\
$\beta_{221}(t)$ & $0.216\,(\phantom{-}0.017)$ & $0.226\,(\phantom{-}0.020)$ & $0.225\,(\phantom{-}0.021)$ & $0.319\,(\phantom{-}0.110)$ & $0.228\,(\phantom{-}0.021)$ & $0.308\,(\phantom{-}0.103)$ & $0.230\,(\phantom{-}0.020)$ & $0.315\,(\phantom{-}0.101)$ \\
$\beta_{222}(t)$ & $0.148\,(\phantom{-}0.012)$ & $0.154\,(\phantom{-}0.013)$ & $0.154\,(\phantom{-}0.013)$ & $0.242\,(\phantom{-}0.137)$ & $0.154\,(\phantom{-}0.012)$ & $0.237\,(\phantom{-}0.112)$ & $0.156\,(\phantom{-}0.013)$ & $0.239\,(\phantom{-}0.127)$ \\
$\beta_{223}(t)$ & $0.027\,(\phantom{-}0.002)$ & $0.029\,(\phantom{-}0.002)$ & $0.029\,(\phantom{-}0.002)$ & $0.150\,(\phantom{-}0.125)$ & $0.029\,(\phantom{-}0.002)$ & $0.143\,(\phantom{-}0.114)$ & $0.029\,(\phantom{-}0.002)$ & $0.147\,(\phantom{-}0.118)$ \\
\addlinespace[2pt]
\midrule
\textit{RCI} & \multicolumn{1}{c}{/} & \multicolumn{1}{c}{/} & 196/200 & 196/200 & 194/200 & 180/200 & 192/200 & 191/200 \\
\textit{IIC} & 100.00\% & 95.53\% & 95.42\% & 67.17\% & 95.44\% & 69.36\% & 95.43\% & 68.63\% \\
Test ACC & \multicolumn{1}{c}{/} & 90.51\% & 90.48\% & 79.60\% & 90.51\% & 80.31\% & 90.53\% & 80.17\% \\
Brier & \multicolumn{1}{c}{/} & 0.068 & 0.069 & 0.188 & 0.068 & 0.182 & 0.068 & 0.184 \\
NLP & \multicolumn{1}{c}{/} & 0.220 & 0.224 & 0.561 & 0.223 & 0.546 & 0.222 & 0.550 \\
$\mathrm{RMSE}_p$ & \multicolumn{1}{c}{/} & 0.034 & 0.042 & 0.341 & 0.041 & 0.331 & 0.041 & 0.334 \\
\bottomrule
\end{tabular}
}

\par\vspace{6pt}
\fontsize{7.8}{9.5}\selectfont
\raggedright
\textit{Note.} Parameter entries are RMSE (bias). Scalar bias is signed;
functional bias is the root integrated squared bias. RCI uses all 200
replications. Parameter errors, IIC, and test metrics use method-specific
fits selecting two subgroups, irrespective of numerical convergence.
\par\vspace{2pt}
The oracle uses known training labels. HLME fixes two groups and the
cosine and sine random-effect functions; unavailable quantities are
marked ``/''. Test prediction uses pretreatment covariates only.
Proposed: $\lambda=0.02$, $|\eta|\leq5$; two-step: $\lambda=0.04$.

\endgroup
\end{minipage}%
}
\end{table}
\clearpage
\begin{table}[p]
\centering
\makebox[\linewidth][c]{%
\begin{minipage}{\linewidth}
\captionsetup{font=small,labelfont=bf,labelsep=simulationbar,justification=raggedright,singlelinecheck=false,name=TABLE}
\caption{Simulation results for $n=800$ (200 replications per method).}
\label{tab:simulation-final-n800}
\begingroup
\fontsize{8.5}{10.2}\selectfont
\setlength{\tabcolsep}{2.3pt}
\renewcommand{\arraystretch}{0.86}
\setlength{\heavyrulewidth}{0.65pt}
\setlength{\lightrulewidth}{0.35pt}
\setlength{\cmidrulewidth}{0.3pt}
\resizebox{\linewidth}{!}{%
\begin{tabular}{@{}lrr@{\hspace{8pt}}rr@{\hspace{8pt}}rr@{\hspace{8pt}}rr@{}}
\toprule
\multirow{2}{*}{Parameter} & \multicolumn{1}{c}{\multirow{2}{*}{\textbf{Oracle}}} & \multicolumn{1}{c}{\multirow{2}{*}{\textbf{HLME} $(G=2)$}} & \multicolumn{2}{c}{\boldmath$G=5$} & \multicolumn{2}{c}{\boldmath$G=8$} & \multicolumn{2}{c}{\boldmath$G=10$} \\
\cmidrule(lr){4-5}\cmidrule(lr){6-7}\cmidrule(l){8-9}
 & & & \multicolumn{1}{c}{\textbf{Proposed}} & \multicolumn{1}{c}{\textbf{Two-step}} & \multicolumn{1}{c}{\textbf{Proposed}} & \multicolumn{1}{c}{\textbf{Two-step}} & \multicolumn{1}{c}{\textbf{Proposed}} & \multicolumn{1}{c}{\textbf{Two-step}} \\
\midrule
$\lambda_{11}$ & $0.067\,(-0.003)$ & $0.063\,(-0.008)$ & $0.068\,(-0.003)$ & $0.102\,(\phantom{-}0.046)$ & $0.068\,(-0.003)$ & $0.102\,(\phantom{-}0.052)$ & $0.068\,(-0.003)$ & $0.106\,(\phantom{-}0.049)$ \\
$\lambda_{12}$ & $0.020\,(-0.003)$ & $0.017\,(-0.001)$ & $0.021\,(-0.003)$ & $0.027\,(\phantom{-}0.011)$ & $0.021\,(-0.003)$ & $0.027\,(\phantom{-}0.011)$ & $0.021\,(-0.003)$ & $0.028\,(\phantom{-}0.012)$ \\
$\lambda_{21}$ & $0.088\,(-0.020)$ & $0.084\,(-0.019)$ & $0.091\,(-0.021)$ & $0.130\,(-0.083)$ & $0.091\,(-0.021)$ & $0.133\,(-0.086)$ & $0.090\,(-0.021)$ & $0.133\,(-0.086)$ \\
$\lambda_{22}$ & $0.024\,(-0.003)$ & $0.020\,(-0.004)$ & $0.025\,(-0.004)$ & $0.035\,(-0.020)$ & $0.025\,(-0.004)$ & $0.034\,(-0.020)$ & $0.025\,(-0.004)$ & $0.036\,(-0.022)$ \\
\addlinespace[1.8pt]
$\omega_{1}$ & \multicolumn{1}{c}{/} & $0.046\,(-0.001)$ & $0.069\,(-0.001)$ & $0.030\,(\phantom{-}0.002)$ & $0.071\,(\phantom{-}0.000)$ & $0.029\,(\phantom{-}0.002)$ & $0.070\,(-0.001)$ & $0.030\,(\phantom{-}0.000)$ \\
$\omega_{2}$ & \multicolumn{1}{c}{/} & $0.046\,(\phantom{-}0.001)$ & $0.069\,(\phantom{-}0.001)$ & $0.030\,(-0.002)$ & $0.071\,(\phantom{-}0.000)$ & $0.029\,(-0.002)$ & $0.070\,(\phantom{-}0.001)$ & $0.030\,(\phantom{-}0.000)$ \\
$\eta_{21}$ & \multicolumn{1}{c}{/} & $0.285\,(-0.067)$ & $0.304\,(-0.071)$ & $1.852\,(\phantom{-}1.846)$ & $0.318\,(-0.082)$ & $1.855\,(\phantom{-}1.849)$ & $0.308\,(-0.085)$ & $1.860\,(\phantom{-}1.854)$ \\
$\eta_{22}$ & \multicolumn{1}{c}{/} & $0.282\,(-0.079)$ & $0.301\,(-0.083)$ & $1.849\,(\phantom{-}1.843)$ & $0.316\,(-0.095)$ & $1.850\,(\phantom{-}1.845)$ & $0.306\,(-0.096)$ & $1.856\,(\phantom{-}1.850)$ \\
$\sigma^2_{1}$ & $0.007\,(-0.003)$ & $0.005\,(-0.003)$ & $0.007\,(-0.003)$ & $0.008\,(-0.004)$ & $0.007\,(-0.003)$ & $0.008\,(-0.004)$ & $0.007\,(-0.003)$ & $0.008\,(-0.004)$ \\
$\sigma^2_{2}$ & $0.007\,(-0.003)$ & $0.005\,(-0.003)$ & $0.007\,(-0.003)$ & $0.009\,(-0.005)$ & $0.007\,(-0.003)$ & $0.009\,(-0.005)$ & $0.007\,(-0.003)$ & $0.009\,(-0.004)$ \\
\addlinespace[1.8pt]
$\phi_{11}(t)$ & $0.043\,(\phantom{-}0.003)$ & \multicolumn{1}{c}{/} & $0.045\,(\phantom{-}0.002)$ & $0.050\,(\phantom{-}0.002)$ & $0.045\,(\phantom{-}0.002)$ & $0.050\,(\phantom{-}0.003)$ & $0.046\,(\phantom{-}0.002)$ & $0.050\,(\phantom{-}0.003)$ \\
$\phi_{12}(t)$ & $0.057\,(\phantom{-}0.003)$ & \multicolumn{1}{c}{/} & $0.060\,(\phantom{-}0.003)$ & $0.065\,(\phantom{-}0.004)$ & $0.059\,(\phantom{-}0.003)$ & $0.065\,(\phantom{-}0.004)$ & $0.060\,(\phantom{-}0.003)$ & $0.065\,(\phantom{-}0.003)$ \\
$\phi_{21}(t)$ & $0.040\,(\phantom{-}0.002)$ & \multicolumn{1}{c}{/} & $0.042\,(\phantom{-}0.003)$ & $0.050\,(\phantom{-}0.007)$ & $0.042\,(\phantom{-}0.003)$ & $0.049\,(\phantom{-}0.006)$ & $0.042\,(\phantom{-}0.003)$ & $0.050\,(\phantom{-}0.008)$ \\
$\phi_{22}(t)$ & $0.052\,(\phantom{-}0.004)$ & \multicolumn{1}{c}{/} & $0.056\,(\phantom{-}0.004)$ & $0.065\,(\phantom{-}0.008)$ & $0.055\,(\phantom{-}0.004)$ & $0.065\,(\phantom{-}0.007)$ & $0.055\,(\phantom{-}0.004)$ & $0.066\,(\phantom{-}0.009)$ \\
\addlinespace[1.8pt]
$\beta_{111}(t)$ & $0.122\,(\phantom{-}0.011)$ & $0.133\,(\phantom{-}0.010)$ & $0.133\,(\phantom{-}0.011)$ & $0.233\,(\phantom{-}0.124)$ & $0.133\,(\phantom{-}0.011)$ & $0.236\,(\phantom{-}0.118)$ & $0.133\,(\phantom{-}0.011)$ & $0.236\,(\phantom{-}0.128)$ \\
$\beta_{112}(t)$ & $0.084\,(\phantom{-}0.005)$ & $0.089\,(\phantom{-}0.005)$ & $0.090\,(\phantom{-}0.006)$ & $0.199\,(\phantom{-}0.137)$ & $0.090\,(\phantom{-}0.006)$ & $0.202\,(\phantom{-}0.144)$ & $0.090\,(\phantom{-}0.006)$ & $0.200\,(\phantom{-}0.139)$ \\
$\beta_{113}(t)$ & $0.017\,(\phantom{-}0.001)$ & $0.017\,(\phantom{-}0.001)$ & $0.018\,(\phantom{-}0.001)$ & $0.154\,(\phantom{-}0.133)$ & $0.018\,(\phantom{-}0.001)$ & $0.154\,(\phantom{-}0.133)$ & $0.018\,(\phantom{-}0.001)$ & $0.156\,(\phantom{-}0.136)$ \\
$\beta_{121}(t)$ & $0.171\,(\phantom{-}0.014)$ & $0.181\,(\phantom{-}0.013)$ & $0.182\,(\phantom{-}0.015)$ & $0.267\,(\phantom{-}0.119)$ & $0.182\,(\phantom{-}0.015)$ & $0.278\,(\phantom{-}0.114)$ & $0.183\,(\phantom{-}0.015)$ & $0.267\,(\phantom{-}0.122)$ \\
$\beta_{122}(t)$ & $0.116\,(\phantom{-}0.008)$ & $0.122\,(\phantom{-}0.008)$ & $0.123\,(\phantom{-}0.009)$ & $0.221\,(\phantom{-}0.141)$ & $0.123\,(\phantom{-}0.009)$ & $0.228\,(\phantom{-}0.149)$ & $0.123\,(\phantom{-}0.009)$ & $0.226\,(\phantom{-}0.143)$ \\
$\beta_{123}(t)$ & $0.024\,(\phantom{-}0.001)$ & $0.025\,(\phantom{-}0.001)$ & $0.025\,(\phantom{-}0.001)$ & $0.155\,(\phantom{-}0.133)$ & $0.025\,(\phantom{-}0.001)$ & $0.155\,(\phantom{-}0.134)$ & $0.025\,(\phantom{-}0.001)$ & $0.157\,(\phantom{-}0.137)$ \\
$\beta_{211}(t)$ & $0.137\,(\phantom{-}0.015)$ & $0.146\,(\phantom{-}0.018)$ & $0.147\,(\phantom{-}0.016)$ & $0.240\,(\phantom{-}0.139)$ & $0.147\,(\phantom{-}0.016)$ & $0.232\,(\phantom{-}0.128)$ & $0.146\,(\phantom{-}0.016)$ & $0.236\,(\phantom{-}0.139)$ \\
$\beta_{212}(t)$ & $0.096\,(\phantom{-}0.010)$ & $0.102\,(\phantom{-}0.010)$ & $0.103\,(\phantom{-}0.010)$ & $0.190\,(\phantom{-}0.128)$ & $0.102\,(\phantom{-}0.010)$ & $0.196\,(\phantom{-}0.137)$ & $0.102\,(\phantom{-}0.010)$ & $0.193\,(\phantom{-}0.129)$ \\
$\beta_{213}(t)$ & $0.017\,(\phantom{-}0.002)$ & $0.018\,(\phantom{-}0.001)$ & $0.018\,(\phantom{-}0.001)$ & $0.152\,(\phantom{-}0.131)$ & $0.018\,(\phantom{-}0.001)$ & $0.152\,(\phantom{-}0.132)$ & $0.018\,(\phantom{-}0.001)$ & $0.153\,(\phantom{-}0.135)$ \\
$\beta_{221}(t)$ & $0.177\,(\phantom{-}0.016)$ & $0.187\,(\phantom{-}0.017)$ & $0.188\,(\phantom{-}0.015)$ & $0.278\,(\phantom{-}0.133)$ & $0.188\,(\phantom{-}0.015)$ & $0.268\,(\phantom{-}0.122)$ & $0.188\,(\phantom{-}0.014)$ & $0.267\,(\phantom{-}0.136)$ \\
$\beta_{222}(t)$ & $0.132\,(\phantom{-}0.006)$ & $0.137\,(\phantom{-}0.006)$ & $0.139\,(\phantom{-}0.006)$ & $0.218\,(\phantom{-}0.132)$ & $0.138\,(\phantom{-}0.006)$ & $0.221\,(\phantom{-}0.142)$ & $0.138\,(\phantom{-}0.005)$ & $0.213\,(\phantom{-}0.131)$ \\
$\beta_{223}(t)$ & $0.024\,(\phantom{-}0.002)$ & $0.024\,(\phantom{-}0.002)$ & $0.025\,(\phantom{-}0.002)$ & $0.153\,(\phantom{-}0.131)$ & $0.025\,(\phantom{-}0.002)$ & $0.153\,(\phantom{-}0.132)$ & $0.025\,(\phantom{-}0.002)$ & $0.154\,(\phantom{-}0.134)$ \\
\addlinespace[2pt]
\midrule
\textit{RCI} & \multicolumn{1}{c}{/} & \multicolumn{1}{c}{/} & 200/200 & 193/200 & 200/200 & 194/200 & 198/200 & 195/200 \\
\textit{IIC} & 100.00\% & 95.70\% & 95.61\% & 65.88\% & 95.61\% & 65.70\% & 95.61\% & 65.08\% \\
Test ACC & \multicolumn{1}{c}{/} & 90.65\% & 90.56\% & 81.78\% & 90.55\% & 81.78\% & 90.59\% & 81.01\% \\
Brier & \multicolumn{1}{c}{/} & 0.067 & 0.068 & 0.191 & 0.068 & 0.192 & 0.068 & 0.194 \\
NLP & \multicolumn{1}{c}{/} & 0.219 & 0.221 & 0.568 & 0.221 & 0.569 & 0.221 & 0.573 \\
$\mathrm{RMSE}_p$ & \multicolumn{1}{c}{/} & 0.028 & 0.034 & 0.346 & 0.034 & 0.347 & 0.034 & 0.350 \\
\bottomrule
\end{tabular}
}

\par\vspace{6pt}
\fontsize{7.8}{9.5}\selectfont
\raggedright
\textit{Note.} Parameter entries are RMSE (bias). Scalar bias is signed;
functional bias is the root integrated squared bias. RCI uses all 200
replications. Parameter errors, IIC, and test metrics use method-specific
fits selecting two subgroups, irrespective of numerical convergence.
\par\vspace{2pt}
The oracle uses known training labels. HLME fixes two groups and the
cosine and sine random-effect functions; unavailable quantities are
marked ``/''. Test prediction uses pretreatment covariates only.
Proposed: $\lambda=0.02$, $|\eta|\leq5$; two-step: $\lambda=0.04$.

\endgroup
\end{minipage}%
}
\end{table}
\clearpage
\end{landscape}
\endgroup

\clearpage
\begingroup
\captionsetup[figure]{font=small,labelfont=bf}
\begin{figure}[!htbp]
\centering
\includegraphics[width=0.95\linewidth]{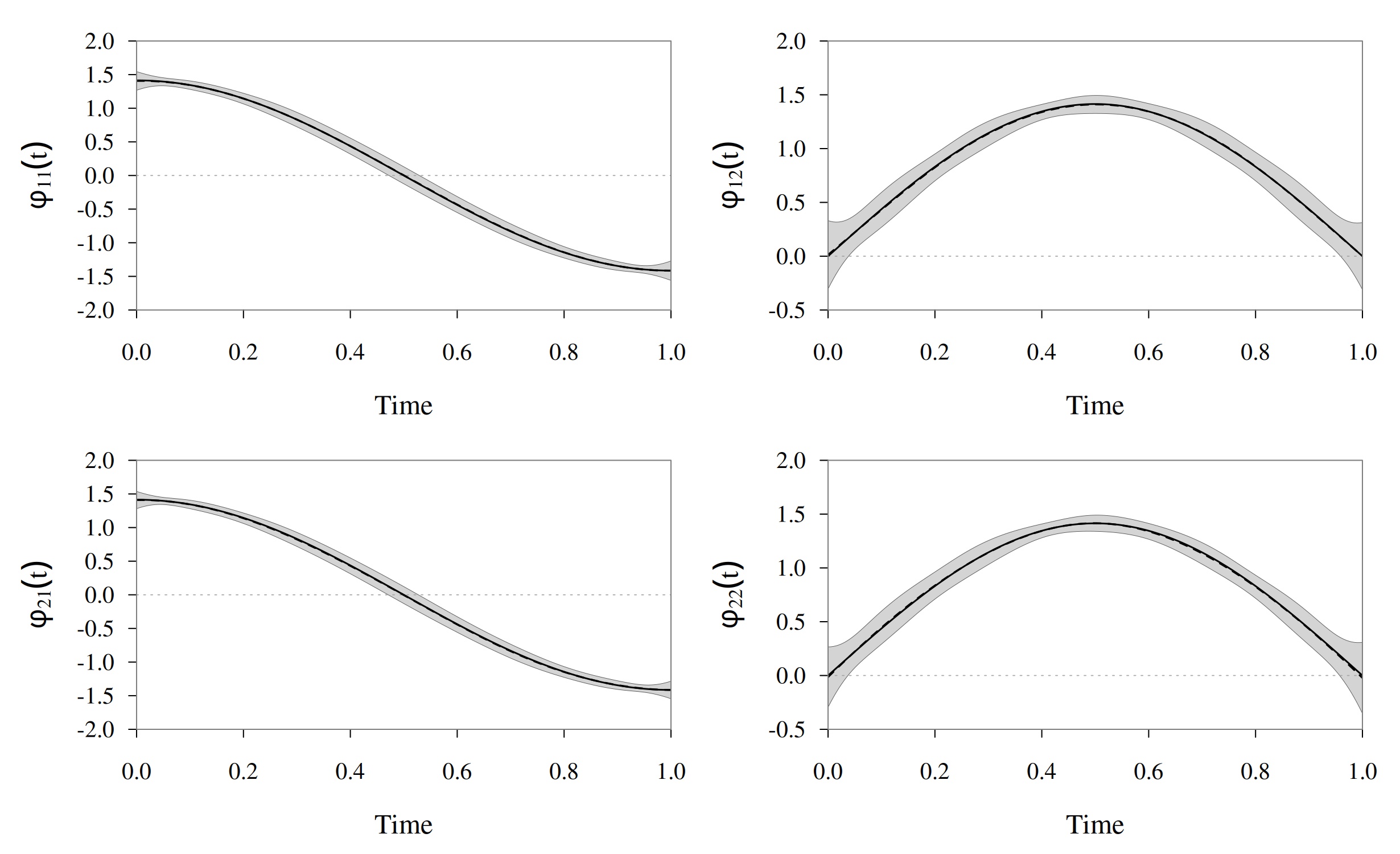}
\caption{Estimated eigenfunctions for $n=600$ and candidate upper bound $\Gmax=5$.
The first row displays $\phi_{11}(t)$ and $\phi_{12}(t)$; the second
displays $\phi_{21}(t)$ and $\phi_{22}(t)$. Solid curves are the true
functions, and dashed curves are averages across replications. Gray areas
are pointwise 95\% Monte Carlo variability bands, constructed as
the average estimate $\pm1.96$ times the pointwise empirical standard
deviation across the 196 replications (of 200) that selected two subgroups.
Curves and band boundaries are evaluated at 200 equally spaced points
on $[0,1]$, including both endpoints. These bands describe sampling
variation across replications.}
\label{fig:simulation-n600-g5-eigen}
\end{figure}

\begin{figure}[!p]
\centering
\includegraphics[width=0.98\linewidth,height=0.82\textheight,keepaspectratio]{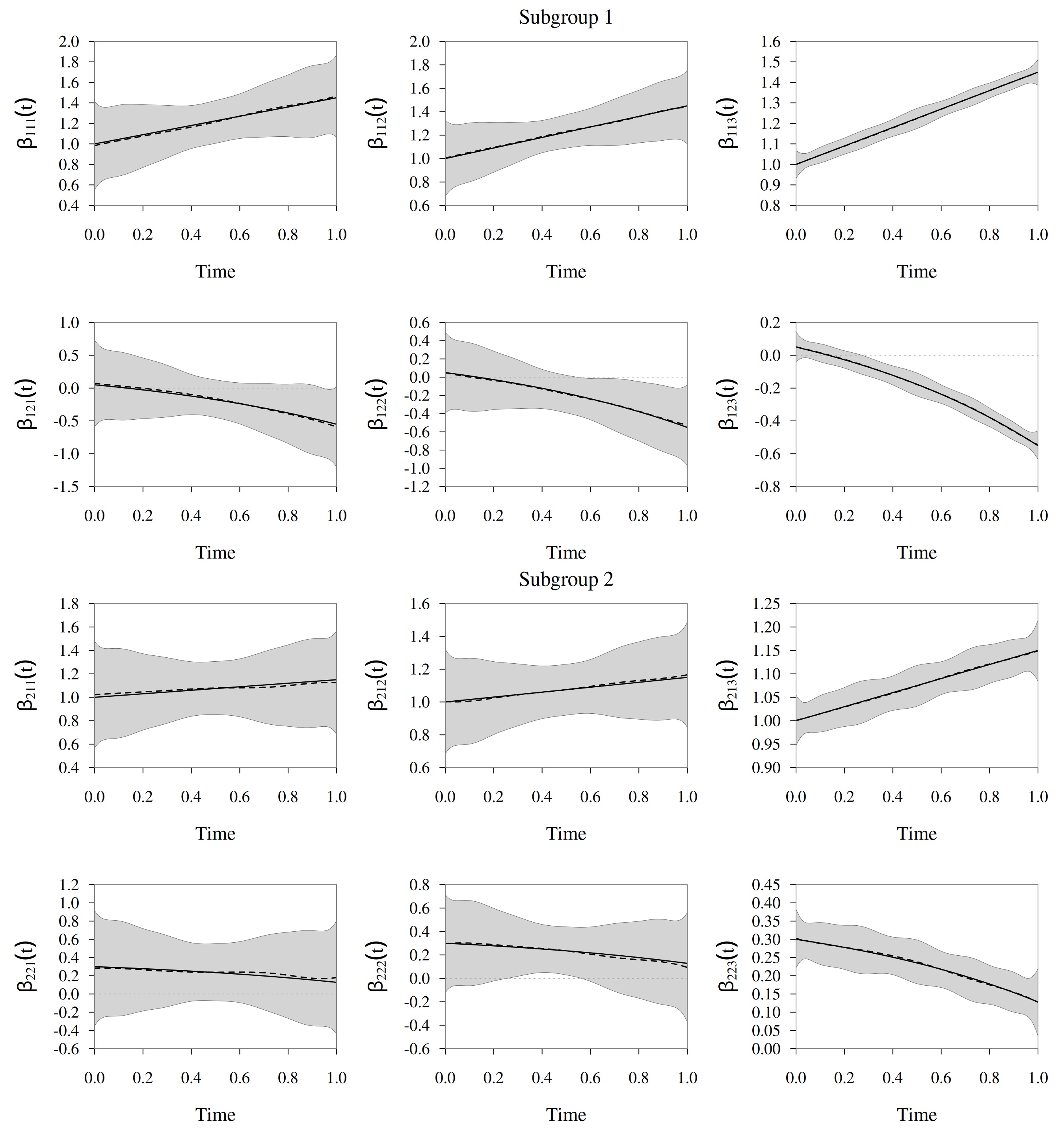}
\caption{Estimated varying coefficients for both subgroups, with $n=600$
and candidate upper bound $\Gmax=5$. The top two rows show subgroup 1, and the bottom two
rows show subgroup 2. Within each subgroup, the first row displays
$\beta_{g11}(t)$, $\beta_{g12}(t)$, and $\beta_{g13}(t)$; the second
shows $\beta_{g21}(t)$, $\beta_{g22}(t)$, and $\beta_{g23}(t)$.
Solid curves, dashed curves, and gray areas denote the true functions,
averages across replications, and pointwise 95\% Monte Carlo variability bands,
respectively, evaluated at 200 equally spaced points on $[0,1]$,
using the same 196 replications and construction as in
Figure~\ref{fig:simulation-n600-g5-eigen}.}
\label{fig:simulation-n600-g5-beta-combined}
\label{fig:simulation-n600-g5-beta1}
\label{fig:simulation-n600-g5-beta2}
\end{figure}
\endgroup

\clearpage

\section{Applications to clinical trial data}\label{sec:application}
\subsection{COMPETE migraine trial}

We illustrate the proposed mixture-of-experts model using the COMPETE
randomized clinical trial of migraine prevention in patients with patent
foramen ovale. COMPETE was a multicenter, randomized, active-controlled
study of antithrombotic agents versus metoprolol
\citep{li2024compete_design}. We compare rivaroxaban with metoprolol,
the active control. The complete-case analysis included 469 participants,
of whom 236 received rivaroxaban and 233 received metoprolol. The
primary trial report evaluated the 12-week responder outcome
\citep{li2026compete}.

The longitudinal outcome was the monthly number of migraine days,
transformed as \(\log(1+Y_{ij})\), and measured at baseline and weeks 4,
8, and 12. Follow-up time was rescaled to \([0,1]\). We set \(D_i=1\)
for rivaroxaban and \(D_i=0\) for metoprolol. The standardized baseline
migraine-attack frequency was included as a prognostic covariate. The
longitudinal model was
\begin{equation*}
\log\{1+\mathrm{MigraineDays}_{ij}\}
=
\sum_{g=1}^{G}\delta_{ig}
\left\{
\beta_{g1}
+D_i\beta_{g2}(t_{ij})
+\beta_{g3}\mathrm{MigraineAttacks}_i
+V_{ig}\phi_g(t_{ij})
+\epsilon_{igj}
\right\}.
\end{equation*}
In this model, \(\beta_{g2}(t)\) is the subgroup-specific treatment-effect
function, approximated by cubic B-splines. The term
\(V_{ig}\phi_g(t)\) represents the leading functional principal
component of within-subject variation.

We used two standardized pretreatment domains from the Migraine-Specific
Quality of Life Questionnaire to model subgroup membership. The Role Restrictive (MSQ-RR) and Emotional Function (MSQ-EF) domains measure
limitations in daily activities and emotional functioning, respectively.
Let
\(\bZ_i=(Z_{i,\mathrm{RR}},Z_{i,\mathrm{EF}})^{\mathrm T}\), without an
intercept. The gating model was
\[
\pi_{ig}
=
\frac{\omega_g\exp(\bZ_i^{\mathrm T}\boldeta_g)}{\sum_{h=1}^{\Gmax}\omega_h
 \exp(\bZ_i^{\mathrm T}\boldeta_h)},
\qquad g=1,\ldots,\Gmax,
\]
where \(\boldeta_1=\bzero\). We initialized \(\Gmax=5\) candidate
components and set \(\lambda=0.01\). The fitted model retained two
subgroups. Table~\ref{tab:migraine_moe_parameters} reports the parameter
estimates and bootstrap summaries.

\begin{table*}[!htbp]
\caption{Parameter estimates for the two identified subgroups in the COMPETE migraine trial.}
\label{tab:migraine_moe_parameters}
\centering
\fontsize{8.5}{10.2}\selectfont
\setlength{\tabcolsep}{3.5pt}
\renewcommand{\arraystretch}{1.12}
\begin{tabular}{@{}lccc@{\hspace{1.4em}}ccc@{}}
\toprule
& \multicolumn{3}{c}{\textbf{Subgroup 1}} & \multicolumn{3}{c}{\textbf{Subgroup 2}} \\
\cmidrule(lr){2-4} \cmidrule(lr){5-7}
\textbf{Parameter} & \textbf{Estimate} & \textbf{SD} & \textbf{95\% CI}
& \textbf{Estimate} & \textbf{SD} & \textbf{95\% CI} \\
\midrule
\multicolumn{7}{@{}l}{\textbf{Longitudinal and covariance parameters}} \\
\addlinespace[1pt]
$\omega_g$       & 0.365 & 0.108 & $(0.191,\ 0.616)$ & 0.635 & 0.108 & $(0.384,\ 0.809)$ \\
$\beta_{g1}$     & 1.988 & 0.071 & $(1.890,\ 2.168)$ & 2.176 & 0.042 & $(2.073,\ 2.217)$ \\
$\beta_{g3}$     & 0.322 & 0.088 & $(0.188,\ 0.533)$ & 0.713 & 0.058 & $(0.606,\ 0.832)$ \\
$\sigma_g^2$     & 0.233 & 0.034 & $(0.147,\ 0.279)$ & 0.029 & 0.007 & $(0.012,\ 0.040)$ \\
$\lambda_g$      & 0.548 & 0.072 & $(0.434,\ 0.736)$ & 0.504 & 0.056 & $(0.413,\ 0.630)$ \\
\addlinespace[3pt]
\multicolumn{7}{@{}l}{\textbf{Gating coefficients relative to subgroup 1}} \\
\cmidrule(lr){1-7}
\textbf{Coefficient} & \multicolumn{2}{c}{\textbf{Estimate}} & \textbf{SD} & \multicolumn{2}{c}{\textbf{95\% CI}} & \\
\midrule
$\eta_{21}$ (MSQ-RR) & \multicolumn{2}{c}{0.446} & 0.232 & \multicolumn{2}{c}{$(0.024,\ 0.936)$} & \\
$\eta_{22}$ (MSQ-EF) & \multicolumn{2}{c}{-0.268} & 0.225 & \multicolumn{2}{c}{$(-0.723,\ 0.157)$} & \\
\bottomrule
\end{tabular}
\par\smallskip
\noindent\fontsize{7.8}{9.5}\selectfont\textit{Note.} Point estimates use the full sample.
Longitudinal and covariance SDs and CIs use 200 nonstratified patient-level warm-start bootstrap replications. Gating SDs and CIs use
200 patient-level bootstrap replications stratified by the original
posterior-modal subgroups (162 and 307 patients); all 200 fits converged.
The gating intervals are bootstrap means $\pm1.96$ bootstrap SDs.
Gating coefficients are log-odds coefficients for subgroup 2 relative
to subgroup 1 per one-SD increase in the standardized MSQ score.
\end{table*}

The fitted global weights were
\(\widehat\omega_1=0.365\) and \(\widehat\omega_2=0.635\). Relative to
subgroup 1, the estimated subgroup 2 gating coefficients were 0.446 for
MSQ-RR and \(-0.268\) for MSQ-EF. Their bootstrap SDs were 0.232 and
0.225, respectively. The corresponding bootstrap-mean-centered normal
95\% intervals were \((0.024,0.936)\) and \((-0.723,0.157)\). Thus,
higher MSQ-RR was associated with greater odds of subgroup 2 membership,
whereas the association with MSQ-EF was less precisely estimated.

\begin{figure}[htbp]
    \centering
    \includegraphics[width=\linewidth]{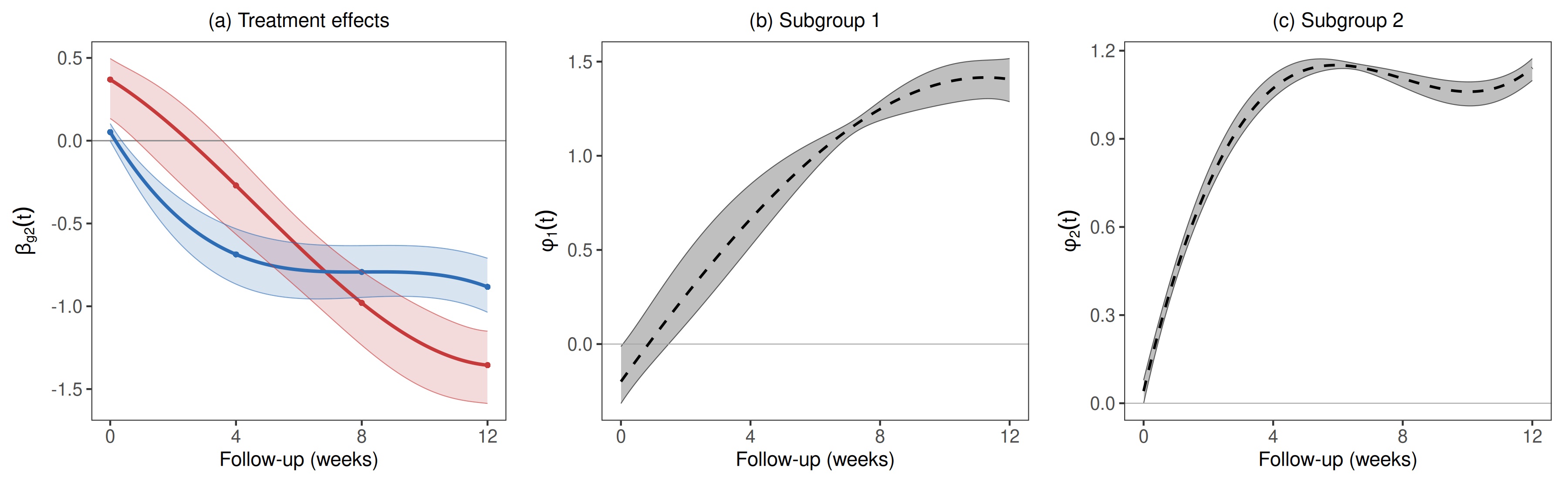}
    \caption{Estimated treatment-effect trajectories and leading eigenfunctions for the
    COMPETE migraine trial. Panel (a) shows \(\beta_{g2}(t)\) for subgroup
    1 (red, delayed-strengthening pattern) and subgroup 2 (blue,
    early-plateau pattern). Panels (b) and (c) show \(\phi_1(t)\) and
    \(\phi_2(t)\), respectively. Curves are fitted estimates; shaded
    regions are pointwise 95\% confidence intervals based on 200
    patient-level warm-start bootstrap resamples. Follow-up is expressed
    in weeks. Negative treatment-effect values indicate a lower
    transformed migraine-day outcome under rivaroxaban than under
    metoprolol.}
    \label{fig:compete_combined}
\end{figure}

Figure~\ref{fig:compete_combined}(a) shows two distinct temporal
treatment-effect patterns. Subgroup 2 exhibited an early-plateau
pattern, with estimated effects of \(-0.68\), \(-0.79\), and \(-0.88\)
at weeks 4, 8, and 12. Subgroup 1 exhibited a delayed-strengthening
pattern, with corresponding estimates of \(-0.27\), \(-0.99\), and
\(-1.37\). The estimated curves crossed during follow-up, showing that
their ordering changed over time.

Because the outcome was measured on the
\(\log(1+\text{migraine days})\) scale, negative estimates indicate a
lower conditional mean under rivaroxaban than under metoprolol. The
estimated heterogeneity therefore concerned both the magnitude and
timing of the treatment contrast. A population-average curve would not
distinguish an early response approaching a plateau from a contrast
that strengthened later.

Figure~\ref{fig:compete_combined}(b,c) shows the estimated leading
eigenfunctions. Their distinct shapes describe subgroup-specific
patterns of within-subject temporal variation beyond the estimated mean
treatment effects.

For a new patient, the fitted gating model gives the subgroup 2 probability
\[
\widehat\pi_2(\bZ_*)
=
\frac{
0.635\exp\{0.446 Z_{*,\mathrm{RR}}-0.268 Z_{*,\mathrm{EF}}\}
}{
0.365+0.635\exp\{0.446 Z_{*,\mathrm{RR}}-0.268 Z_{*,\mathrm{EF}}\}
}.
\]
The scores \(Z_{*,\mathrm{RR}}\) and \(Z_{*,\mathrm{EF}}\) are
standardized using the analysis-sample means and standard deviations,
and \(\widehat\pi_1(\bZ_*)=1-\widehat\pi_2(\bZ_*)\). The subgroup with the larger probability gives a hard prediction,
whereas retaining both probabilities quantifies assignment uncertainty.

\clearpage
\subsection{SANAD epilepsy study}\label{sec:sanad}

The Standard and New Antiepileptic Drugs (SANAD) study was a randomized
trial of standard and newer antiepileptic drugs
\citep{marson2007randomised,marson2007sanad}. Following
\citet{wei2023}, we analyzed patients receiving lamotrigine (LTG) or
carbamazepine (CBZ). The outcome was the log-transformed Liverpool
Adverse Events Profile (AEP), for which higher scores indicate a greater
burden of adverse effects.  The complete-case
dataset contained 1,720 observations from 540 patients, including 262
assigned to LTG and 278 assigned to CBZ. Observation times were divided
by the maximum observed follow-up time, giving \(t_{ij}\in[0,1]\).

For each subject, the first recorded anxiety (ANX) and depression (DEP)
scores were standardized and used as subject-level predictors. The
model included constant prognostic coefficients and a time-varying
treatment effect. Conditional on \(G_i=g\), the fitted longitudinal
model was
\[
\log(\mathrm{AEP}_{ij})
=
\beta_{g1}+D_i\beta_{g2}(t_{ij})
+\beta_{g3}\mathrm{ANX}_i+\beta_{g4}\mathrm{DEP}_i
+V_{ig}\phi_g(t_{ij})+\epsilon_{igj},
\]
where \(D_i=1\) for LTG and \(D_i=0\) for CBZ. The same ANX and DEP
scores entered the gating model in \eqref{eq:gating}. We used cubic
B-splines with interior knots at 0.153 and 0.312 and set \(K=1\).
Starting from \(\Gmax=5\) with \(\lambda=0.03\), the algorithm
converged to two subgroups after 25 iterations.

Table~\ref{tab:sanad-parameters} summarizes the fitted parameters. The
global weights were \(\widehat\omega_1=0.292\) and
\(\widehat\omega_2=0.708\). These are membership probabilities at the
centered predictor profile, not marginal subgroup proportions. ANX and
DEP had positive associations with log-AEP in both subgroups, with
larger estimates in subgroup 2. The estimated latent-process and
measurement-error variances were also larger in subgroup 2.

Of 200 subject-level bootstrap replications, 186 selected two subgroups and
were used for interval estimation; seven of these did not meet the
numerical convergence criterion. Scalar intervals used a normal
approximation, whereas function intervals used pointwise percentiles.
Before summarization, subgroup labels were aligned with the fitted model
and gating coefficients were recentered to the same reference subgroup.

\begin{table}[!htbp]
\centering
\caption{Parameter estimates for the two identified subgroups in the SANAD epilepsy study.}
\label{tab:sanad-parameters}
\fontsize{8.5}{10.2}\selectfont
\setlength{\tabcolsep}{4pt}
\renewcommand{\arraystretch}{1.12}
\begin{tabular}{@{}lccc@{\hspace{1.2em}}ccc@{}}
\toprule
& \multicolumn{3}{c}{\textbf{Subgroup 1}} & \multicolumn{3}{c}{\textbf{Subgroup 2}} \\
\cmidrule(lr){2-4} \cmidrule(lr){5-7}
Parameter & Estimate & SD & 95\% CI & Estimate & SD & 95\% CI \\
\midrule
$\omega_g$ & 0.292 & 0.105 & $(0.087,\ 0.497)$ & 0.708 & 0.105 & $(0.503,\ 0.913)$ \\
$\beta_{g1}$ & 3.849 & 0.043 & $(3.766,\ 3.933)$ & 3.641 & 0.041 & $(3.561,\ 3.721)$ \\
$\beta_{g3}$ (ANX) & 0.049 & 0.020 & $(0.010,\ 0.088)$ & 0.127 & 0.021 & $(0.086,\ 0.169)$ \\
$\beta_{g4}$ (DEP) & 0.045 & 0.014 & $(0.018,\ 0.073)$ & 0.097 & 0.022 & $(0.055,\ 0.139)$ \\
$\sigma_g^2$ & 0.006 & 0.003 & $(0.000,\ 0.012)$ & 0.025 & 0.003 & $(0.018,\ 0.031)$ \\
$\lambda_g$ & 0.012 & 0.006 & $(0.001,\ 0.023)$ & 0.043 & 0.007 & $(0.029,\ 0.056)$ \\
\midrule
\multicolumn{7}{@{}l}{\textbf{Gating coefficients: subgroup 2 relative to subgroup 1}} \\
\cmidrule(lr){1-7}
Coefficient & \multicolumn{2}{c}{Estimate} & SD & \multicolumn{2}{c}{95\% CI} & \\
\midrule
$\eta_{21}$ (ANX) & \multicolumn{2}{c}{-0.993} & 0.473 & \multicolumn{2}{c}{$(-1.920,\ -0.066)$} & \\
$\eta_{22}$ (DEP) & \multicolumn{2}{c}{-0.433} & 0.330 & \multicolumn{2}{c}{$(-1.079,\ 0.213)$} & \\
\bottomrule
\end{tabular}
\par\smallskip
\begin{minipage}{\linewidth}\fontsize{7.8}{9.5}\selectfont
\textit{Note.} ANX and DEP are standardized first-recorded subject-level scores.
SDs and normal-approximation 95\% CIs use the 186 bootstrap replications (of 200) that selected two subgroups; 179 met the numerical convergence
criterion; the other seven fits are retained. Gating coefficients
are recentered to subgroup 1 after label alignment.
\end{minipage}
\end{table}

\begin{figure}[!htbp]
\centering
\includegraphics[width=\linewidth]{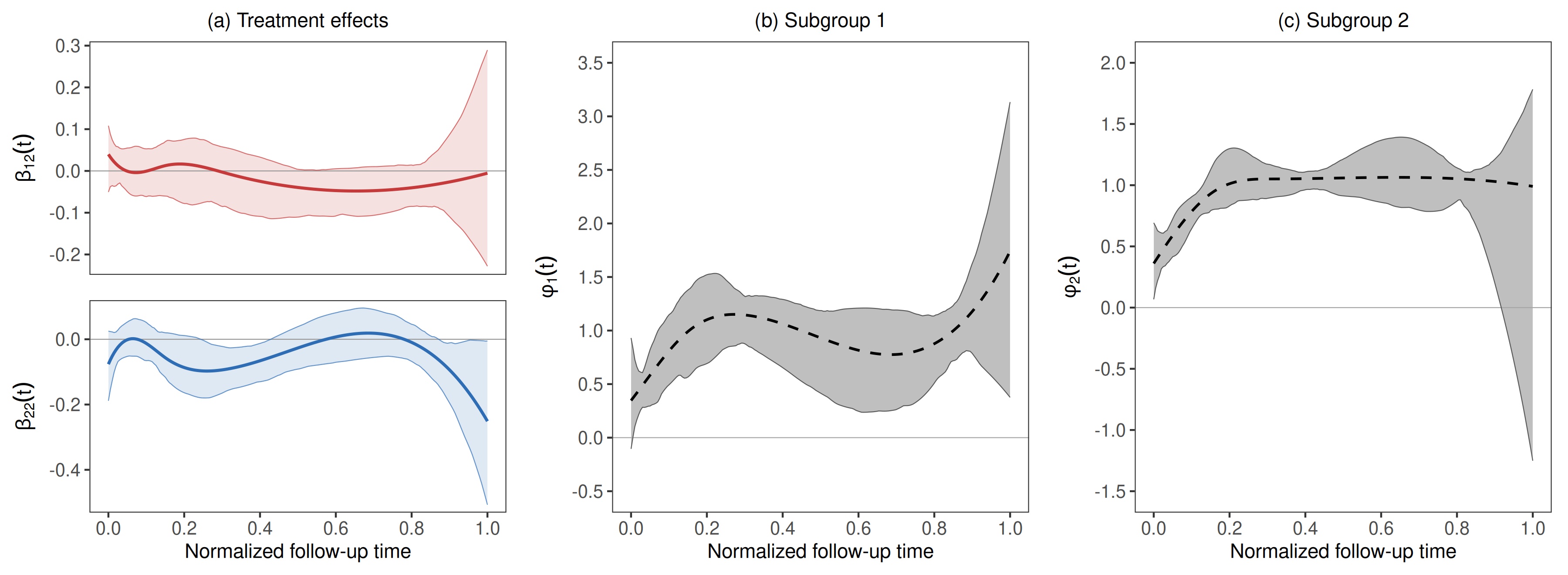}
\caption{Estimated treatment-effect trajectories and leading eigenfunctions for the SANAD epilepsy study. Panel (a) displays \(\beta_{12}(t)\) (upper, red) and
\(\beta_{22}(t)\) (lower, blue) with separate vertical scales.
Panels (b) and (c) display \(\phi_1(t)\) and \(\phi_2(t)\).
Curves are fitted estimates, and shaded regions are pointwise 95\% bootstrap percentile intervals based on the 186
replications (of 200) that selected two subgroups. Eigenfunction signs are aligned to the fitted
functions. Time is normalized to \([0,1]\).}
\label{fig:sanad-combined}
\end{figure}

Figure~\ref{fig:sanad-combined}(a) shows the subgroup-specific treatment
contrasts. The estimated \(\beta_{12}(t)\) remained close to zero, and
its pointwise interval included zero throughout follow-up. In subgroup
2, the estimated contrast was negative during the middle of follow-up,
returned toward zero, and decreased again near the end. Its pointwise
interval lay below zero over approximately \([0.25,0.43]\) and
\([0.87,1.00]\) on the normalized time scale. Negative contrasts
indicate a lower conditional mean log-AEP under LTG than under CBZ. The
late estimates were less precise, as indicated by the widening
intervals. Panels (b) and (c) show the corresponding eigenfunctions.
Subgroup 1 had a late upswing, whereas subgroup 2 was comparatively
stable during most of the later follow-up.

The gating estimates for subgroup 2 relative to subgroup 1 were
\(\widehat\eta_{21}=-0.993\) for ANX and
\(\widehat\eta_{22}=-0.433\) for DEP. The reference-aligned 95\%
intervals were \((-1.920,-0.066)\) and \((-1.079,0.213)\), respectively.
Higher ANX was therefore associated with lower odds of subgroup 2
membership. The association with DEP was less precisely estimated.
Although the fitted gating model uses these two scores alone, their pretreatment
availability has not been established. Using the fitted model at treatment assignment would require confirming
that both measurements precede treatment.

\clearpage
\section{Discussion}

In this article, we proposed a data-adaptive semiparametric
mixture-of-experts model that unifies longitudinal treatment-effect
subgroup discovery and pretreatment membership prediction. Unlike the
two-step strategy, the pretreatment membership model enters the
observed-data likelihood and directly contributes to subgroup discovery.
Separate global component weights and covariate-dependent gating
coefficients allow redundant components to be removed while retaining
patient-specific membership probabilities. The model also accommodates
subgroup-specific treatment-effect trajectories and covariance
functions. Under regularity conditions, it consistently selects the
subgroup number, with established convergence rates and asymptotic
normality. Across candidate upper bounds, simulations showed reliable
subgroup-number selection and generally more accurate recovery of the
longitudinal components and membership probabilities than the two-step
procedure. These comparisons were conditional on selecting two
subgroups; the correctly specified HLME benchmark, supplied with the
true subgroup number and covariance shapes, performed better for some
parameters.

The applications illustrate the value of studying temporal
treatment-effect heterogeneity together with pretreatment membership
prediction. In the SANAD study, the comparison between lamotrigine and
carbamazepine was close to zero in one subgroup but favored lamotrigine
during specific periods in the other. In the COMPETE trial, the
analysis distinguished an early-plateau pattern from a
delayed-strengthening pattern in the comparison of rivaroxaban with
metoprolol. Such temporal differences would be obscured by a
population-average trajectory. The COMPETE analysis also translated
pretreatment characteristics into membership probabilities for new
patients. Retaining these probabilities instead of a hard assignment
preserves uncertainty when subgroup separation is incomplete. These
clinical findings remain exploratory and require independent
validation.

Several limitations remain. Because the latent subgroups are defined
by the full longitudinal distribution, they may differ in prognosis or
within-subject covariance as well as in treatment response and need not
represent distinct treatment-sensitivity phenotypes. The working model
assumes Gaussian component outcomes, a finite-rank covariance
representation, smooth coefficient functions, and a multinomial-logit
gating model. The penalized mixture objective is nonconvex, and both
applications used complete-case data. Future work could accommodate
sparse or nonlinear gating effects, non-Gaussian or multivariate
longitudinal outcomes, informative dropout, and subgroup-selection
uncertainty. External validation is essential because prediction
requires the learned subgroup structure to transport to the target
population. Finally, accurate membership prediction does not establish
an optimal treatment rule; decision-oriented methods must also account
for membership uncertainty and the consequences of incorrect treatment
assignment.


\providecommand{\bmsection}{\section}
\bibliographystyle{plainnat}
\bibliography{references,manu_arxiv_model_references}

\end{document}